\documentclass[12pt]{iopart}
\usepackage{iopams} 
\usepackage{graphicx}
\usepackage{dcolumn}
\usepackage{bm}
\usepackage{braket}                             
\usepackage{amsfonts}
\usepackage{bbold}                              
\usepackage{subfigure}                          
\usepackage{hyperref}                           
\usepackage[dvipsnames]{xcolor}                 
\usepackage{multirow}                          
\usepackage{enumitem}                           
\newcommand{\argmin}{\mathop{\mathrm{arg\,min}}}  
\usepackage{algorithm}  
\usepackage[noend]{algpseudocode}   
\algrenewcommand\alglinenumber[1]{\scriptsize #1:}
\algnewcommand{\AND}{\textbf{and}}
\algnewcommand{\OR}{\textbf{or}}
\algnewcommand{\NOT}{\textbf{not}}
\usepackage{orcidlink}  

\begin{document}

\title[]{Experimental demonstration of improved quantum optimization with linear Ising penalties}

\author{Puya~Mirkarimi\,\orcidlink{0000-0001-5835-2592}}
    \address{Department of Physics, Durham University, Durham DH1 3LE, United Kingdom}
    \ead{puya.mirkarimi@durham.ac.uk}
\author{David~C.~Hoyle\,\orcidlink{0000-0003-3483-5885}}
    \address{dunnhumby, 184 Shepherds Bush Road, London W6 7NL, United Kingdom}
\author{Ross~Williams\,\orcidlink{0000-0001-9347-0922}}
    \address{dunnhumby, 184 Shepherds Bush Road, London W6 7NL, United Kingdom}
\author{Nicholas~Chancellor\,\orcidlink{0000-0002-1293-0761}}
    \address{Department of Physics, Durham University, Durham DH1 3LE, United Kingdom}
    \vspace{5pt}
    \address{School of Computing, Newcastle University, 1 Science Square, Newcastle upon Tyne NE4 5TG, United Kingdom}
    \ead{nicholas.chancellor@gmail.com}

\vspace{10pt}
\begin{indented}
\item[]16 December 2024
\end{indented}

\begin{abstract}
The standard approach to encoding constraints in quantum optimization is the quadratic penalty method. Quadratic penalties introduce additional couplings and energy scales, which can be detrimental to the performance of a quantum optimizer. In quantum annealing experiments performed on a D-Wave Advantage, we explore an alternative penalty method that only involves linear Ising terms and apply it to a customer data science problem. Our findings support our hypothesis that the linear Ising penalty method should improve the performance of quantum optimization compared to using the quadratic penalty method due to its more efficient use of physical resources. Although the linear Ising penalty method is not guaranteed to exactly implement the desired constraint in all cases, it is able to do so for the majority of problem instances we consider. For problems with many constraints, where making all penalties linear is unlikely to be feasible, we investigate strategies for combining linear Ising penalties with quadratic penalties to satisfy constraints for which the linear method is not well-suited. We find that this strategy is most effective when the penalties that contribute most to limiting the dynamic range are removed.
\end{abstract}

\maketitle

\section{\label{sec:introduction}Introduction}

In recent years, noisy intermediate-scale quantum devices have become available on cloud platforms~\cite{Preskill2018}, enabling small-scale experimental tests of quantum algorithms. One area of particular interest is combinatorial optimization~\cite{Au-Yeung2023}, in which the goal is to minimise an objective function that can take a finite but very large set of possible solutions. Applications of quantum optimization have been explored in a variety of fields including finance~\cite{Orus2019, Venturelli2019}, molecular biology~\cite{Fox2021}, material design~\cite{Kitai2020}, and air traffic management~\cite{Stollenwerk2020}. In this work, we consider an application in customer data science and perform experiments on a quantum annealer developed by D-Wave Systems~\cite{Johnson2011}. For a review of quantum annealing for industrial applications, see Ref.~\cite{Yarkoni2021}.

Combinatorial optimization problems that arise in industry are often highly constrained~\cite{Yarkoni2021}. In commerce and retail settings, constraints can number in the tens to hundreds and result from both strategic and operational considerations. The large number of constraints influences how a researcher can tackle an optimization problem in practice. Therefore, it is important to study quantum optimization problems that include multiple constraints, since if quantum optimization is to find real value in a commercial setting, it must be able to handle and thrive in problems significantly affected by or even dominated by constraints. In quantum annealing (QA)~\cite{Kadowaki1998, Brooke1999, Farhi2001}, constraints are typically incorporated into the objective function by adding terms that penalise solutions that do not satisfy the constraints. Typically, these terms are quadratic in the input variables and are therefore called quadratic penalty functions. Devices that implement QA have certain physical limitations, such as the number of qubits, the number of two-qubit interactions that can be realised, and the maximum energy scale over which an interaction can take~\cite{Advantage6-3}. Quadratic penalty functions are physically intensive with respect to these resources, hindering the performance of a quantum algorithm when reconciled with the hardware's physical limitations.

While the number of interactions and the structure of the interactivity graph are not generally considered to be important limiting factors when solving optimization problems through traditional methods, they do present a significant challenge in the analog setting of quantum annealing, particularly on current devices where problems need to be minor embedded onto quasi-planar graphs~\cite{Choi2008}. For this reason, new methods for encoding problems need to be considered in this setting, such as domain-wall encoding~\cite{Chancellor2019}, which has been shown to provide a larger performance enhancement than a re-engineered hardware graph~\cite{Chen2021} and make the dynamics more favourable~\cite{Berwald2023}. While domain-wall encoding can only be used efficiently to transform one-hot constraints, the method we propose here can be applied to any kind of linear constraint, including inequality constraints.

In this work, we consider a linear Ising penalty method as an alternative to the quadratic penalty method for QA. While it is not guaranteed that this method can exactly implement the desired constraint in all cases, it makes more efficient use of physical resources than the quadratic penalty method. Notably, the linear Ising penalty method does not introduce any additional two-qubit interactions and often requires smaller energy scales to implement. This method has been suggested in previous studies~\cite{Venturelli2019, Ohzeki2020, Willsch2020a}. Ref.~\cite{theorypaper} explores the viability of the linear Ising penalty method through a theoretical and numerical analysis. Here, we provide complementary results from experiments on real quantum hardware. Our findings indicate that for some problems, the linear Ising penalty method results in better performance of quantum optimization compared to the quadratic penalty method. For problems with multiple constraints, we find that a strategy of using linear Ising penalties where possible and quadratic penalties for the remaining constraints has the potential to work well in practice. By using a combination of the two penalty methods, we can gain some of the advantages of the linear Ising penalty method while still satisfying constraints that it cannot implement exactly.

This paper is structured as follows.
In Sec.~\ref{sec:background}, we summarise prior work, introduce the customer data science problems that we consider, and outline the linear Ising penalty method. In Sec.~\ref{sec:numerical_and_experimental_methods}, we describe the numerical and experimental methods used in this work. In Sec.~\ref{sec:minor_embedding_improvements}, we analyse the minor embedding costs of the two penalty methods and how they scale with problem size. We present the results of our experiments on a D-Wave device in Sec.~\ref{sec:dwave_results} and compare the performance of the two penalty methods. Finally, we summarise our work and give some concluding remarks in Sec.~\ref{sec:conclusions}.

\section{\label{sec:background}Background}

In this section, we provide a background on prior work and introduce the key concepts in this study. Sec.~\ref{sec:promotion_cannibalization_problem} describes the customer data science problems that we base our experiments on. Sec.~\ref{sec:penalty_methods} describes how constraints are encoded with the quadratic and linear Ising penalty methods.

\subsection{\label{sec:promotion_cannibalization_problem}Promotion cannibalization problem}

Quantum optimizers, such as the implementation of QA on D-Wave quantum annealers, can be used to find ground states of the Ising model. For local fields $\textbf{h} \in \mathbb{R}^n$ and couplings $\mathbf{J} \in \mathbb{R}^{n \times n}$, the Ising model is described by the Hamiltonian
\begin{equation}
    H_P = \sum_{i=1}^{n} h_i \sigma_i^z + \sum_{i=1}^{n-1} \sum_{j=i+1}^{n} J_{i,j} \sigma_i^z \sigma_j^z,
    \label{eq:ising_hamiltonian}
\end{equation}
where $\sigma_i^z = \mathbb{1}^{\otimes i - 1} \otimes \sigma_z \otimes \mathbb{1}^{\otimes n - i}$ is the Pauli operator $\sigma_z$ acting on qubit $i$ and identities acting on all other qubits. Note that $\mathbf{J}$ is a strictly upper triangular matrix.

The Pauli matrices in the Ising Hamiltonian can be mapped to binary variables $\mathbf{x} \in \{0, 1\}^n$ through the mapping $\sigma_i^z \mapsto 1 - 2 x_i$. Therefore, finding a ground state of $H_P$ is equivalent to the quadratic unconstrained binary optimization (QUBO) problem
\begin{eqnarray}
    \label{eq:QUBO_objective_function}
    \mathrm{find}:~~~&& \argmin_\mathbf{x} f(\mathbf{x}) = \sum_{i=1}^{n} a_i x_i + \sum_{i=1}^{n-1} \sum_{j=i+1}^{n} b_{i,j} x_i x_j.
\end{eqnarray}
The real-valued coefficients $\mathbf{a}$ and $\mathbf{b}$ in the objective function $f(\mathbf{x})$ are related to $\textbf{h}$ and $\mathbf{J}$ through the equations
\begin{equation}
    J_{i,j} = \frac{b_{i,j}}{4}
    \label{eq:ising_couplings_qubo_relation}
\end{equation}
and
\begin{equation}
    h_{i} = - \frac{a_i}{2} - \frac{1}{4} \sum_{j=1, j \neq i}^n b_{i,j}.
    \label{eq:ising_field_strengths_qubo_relation}
\end{equation}
In other studies, the QUBO problem is often expressed in terms of a single upper triangular matrix $Q \in \mathbb{R}^{n \times n}$, where $Q_{i,j} = b_{i,j}~\forall j \neq i$ and $Q_{i,i} = a_i$.

In this work, we consider two simplified forms of a customer data science problem that is faced by retailers when planning product price reduction promotions. The goal of a promotion is to generate additional revenue from sales of a product. Often, these additional sales come at the partial expense of other products' sales. This phenomenon, where a product takes sales from other products instead of generating new sales, is called cannibalization~\cite{Meredith2001, Aguilar-Palacios2021}, and we will refer to cannibalization arising from promoting a product as \textit{promotion cannibalization}. When promoting two similar products concurrently, the overall bilateral cannibalization can result in minimal new sales and possibly even a net reduction in revenue. Consequently, one goal of promotion planning for retailers is to minimize the overall revenue loss due to clashing concurrent promotions. We can model this by considering only promotion cannibalization between pairs of products that are promoted at the same time, and using a matrix $C$, where the matrix element $C_{i,j}$ represents the average amount of loss of revenue from sales of product $i$ due to a promotion of product $j$ when both products are promoted at the same time. For this study, we will assume that $C_{i,j} \geq 0~\forall i, j$.

An example problem that we use in this work is to find a promotion plan for one fiscal year that minimises the total amount of cannibalization between pairs of products that are promoted in the same fiscal quarter. The promotion plan indicates which products are promoted in each quarter, and it must satisfy various constraints that are imposed by the retailer. We consider three sets of constraints:
\begin{enumerate}[label=C\arabic*., ref=C\arabic*]
    \item Each quarter must have $A$ products promoted.
    \label{constraint:C1}
    \item Each product must be promoted between $B_\mathrm{min}$ and $B_\mathrm{max}$ times by the end of the year.
    \label{constraint:C2}
    \item The same product cannot be promoted in two consecutive quarters.
    \label{constraint:C3}
\end{enumerate}
In practice, there may be many more constraints that a retailer would want to implement.

This problem can be expressed as the constrained binary quadratic programming problem~\cite{nocedal1999numerical, VanThoai2013}
\begin{eqnarray}
    \label{eq:four_quarter_promotion_cannibalization_qubo_objective_function}
    \mathrm{find}:~~~&& \argmin_\mathbf{x} f(\mathbf{x}) = \sum_{q=1}^4 \sum_{j=1}^{n_p} \sum_{i=1}^{n_p} \lambda_q C_{i,j} x_{i,q} x_{j,q} \\
    \label{eq:constraint_C1}
    \mathrm{subject~to}:~~~&& \sum_{i=1}^{n_p} x_{i,q} = A~~\forall q, \\
    \label{eq:constraint_C2}
    && B_\mathrm{min} \leq \sum_{q=1}^4 x_{i,q} \leq B_\mathrm{max}~~\forall i, \\
    \label{eq:constraint_C3}
    && x_{i,q} + x_{i,q+1} \leq 1~~\forall q \leq 3~~\forall i.
\end{eqnarray}
Here, the binary variable $x_{i,q} = 1$ ($= 0$) if product $i$ is (not) promoted during fiscal quarter $q$, $n_p$ is the total number of products, and $\lambda_q$ is a seasonal scale factor representing the expected changes in total sales between quarters. Eqs.~\ref{eq:constraint_C1}, \ref{eq:constraint_C2}, and \ref{eq:constraint_C3} correspond to the sets of constraints~\ref{constraint:C1}, \ref{constraint:C2}, and \ref{constraint:C3} respectively. Comparing Eq.~\ref{eq:four_quarter_promotion_cannibalization_qubo_objective_function} with Eq.~\ref{eq:QUBO_objective_function}, we note that the quadratic coefficients in the objective function are equal to $\lambda_q (C_{i,j} + C_{j,i})$. Hence, while $C$ is generally asymmetric, it can be assumed to be symmetric in the context of this problem.

The constraints~\ref{constraint:C1} are linear equality constraints, which take the general form
\begin{equation}
    \sum_{i=1}^n \mu_i x_i = c,
    \label{eq:linear_equality_constraint}
\end{equation}
with some coefficients $\boldsymbol{\mu} \in \mathbb{R}^n$ and constraint value $c \in \mathbb{R}$. The constraints~\ref{constraint:C2} and~\ref{constraint:C3} are linear \textit{inequality} constraints of the form
\begin{equation}
    d_\mathrm{min} \leq \sum_{i=1}^n \nu_i x_i \leq d_\mathrm{max},
    \label{eq:linear_inequality_constraint}
\end{equation}
with some coefficients $\boldsymbol{\nu} \in \mathbb{R}^n$ and constraint values $d_\mathrm{min}, d_\mathrm{max} \in \mathbb{R}$.

We also consider a simpler formulation of the promotion cannibalization problem, in which we are only concerned with the promotion plan for a single quarter. In this case, the relevant quadratic programming problem is
\begin{eqnarray}
    \label{eq:single_quarter_promotion_cannibalization_qubo_objective_function}
    \mathrm{find}:~~~&& \argmin_\mathbf{x} f(\mathbf{x}) = \sum_{j=1}^{n_p} \sum_{i=1}^{n_p} C_{i,j} x_i x_j \\
    \label{eq:constraint_C1_single_quarter}
    \mathrm{subject~to}:~~~&& \sum_{i=1}^{n_p} x_i = A.
\end{eqnarray}
Note that this problem only has one constraint. To distinguish between the two variations of the promotion cannibalization problem, we will refer to Eqs.~\ref{eq:four_quarter_promotion_cannibalization_qubo_objective_function}--\ref{eq:constraint_C3} as the \textit{four-quarter} problem and Eqs.~\ref{eq:single_quarter_promotion_cannibalization_qubo_objective_function}--\ref{eq:constraint_C1_single_quarter} as the \textit{single-quarter} problem.

As a metric to assess the quality of a solution $\mathbf{x}$, we use the approximation ratio
\begin{equation}
    R = 1- \frac{f(\mathbf{x}) - f_\mathrm{min}}{f_\mathrm{max} - f_\mathrm{min}},
\end{equation}
where $f_\mathrm{min}$ and $f_\mathrm{max}$ are the minimum and maximum objective values of the constrained problem respectively. When $\mathbf{x}$ is feasible, i.e.\ satisfies all constraints, $R$ ranges from $0$ in the worst case ($f(\mathbf{x}) = f_\mathrm{max}$) to $1$ for optimal solutions ($f(\mathbf{x}) = f_\mathrm{min}$). If an optimizer is run many times, we refer to the fraction of sampled solutions that are optimal as $S$. Similarly, we refer to the fraction of samples that are feasible as $F$.

\subsection{\label{sec:penalty_methods}Penalty methods for encoding constraints}

The promotion cannibalization problems expressed in the forms of Eqs.~\ref{eq:four_quarter_promotion_cannibalization_qubo_objective_function}--\ref{eq:constraint_C3} and Eqs.~\ref{eq:single_quarter_promotion_cannibalization_qubo_objective_function}--\ref{eq:constraint_C1_single_quarter} are not QUBO problems because they include constraints, which cannot be natively expressed in QA. Various methods for encoding constraints in QA have been proposed. The most common approach on D-Wave quantum annealers is to incorporate constraints into the objective function using the penalty method, where for each constraint, a penalty function $P(\textbf{x})$ is added to $f(\mathbf{x})$. The purpose of the penalty method is to raise the value of $f(\mathbf{x})$ for solutions that don't satisfy all constraints (which are called infeasible solutions) so that solutions that minimise the penalised objective function are feasible.

To encode equality constraints of the form given in Eq.~\ref{eq:linear_equality_constraint}, the quadratic penalty function
\begin{equation}
    P(\mathbf{x}) = \alpha_2 \left(\sum_{i=1}^n \mu_i x_i - c \right)^2
    \label{eq:quadratic_penalty_function}
\end{equation}
is typically used in QA. For a large enough value of the penalty strength $\alpha_2$, adding $P(\textbf{x})$ to an objective function will make it satisfy Eq.~\ref{eq:linear_equality_constraint}. Aside from the ability to scale the penalty function with $\alpha_2$, this penalty function has two desirable properties that hold for all $\alpha_2 > 0$:
\begin{enumerate}
    \item $P(\textbf{x}) = 0$ if $\textbf{x}$ is feasible.
    \label{item:penalty_function_property_1}
    \item $P(\textbf{x}) > 0$ if $\textbf{x}$ is infeasible.
    \label{item:penalty_function_property_2}
\end{enumerate}

The quadratic penalty method can impose severe limitations in QA. Expanding out the brackets of the quadratic penalty in Eq.~\ref{eq:quadratic_penalty_function}, we find quadratic terms with nonzero coefficients for all pairs of variables in the constraint. This leads to an Ising Hamiltonian $H_P$ with all-to-all couplings between the qubits associated with the constraint. Many quantum devices, including the D-Wave quantum annealers, have limited qubit connectivity. Because of this, most Ising Hamiltonians of interest cannot be directly mapped to the hardware. On D-Wave devices, the lack of connectivity is resolved by representing each variable by a logical qubit formed of a chain of ferromagnetically coupled physical qubits through a mapping called minor embedding~\cite{Choi2008, Yarkoni2021}. After minor embedding, the necessary couplings are available to the logical qubits at the cost of a larger number of physical qubits being used. Typically, $\mathcal{O}(n^2)$ physical qubits are required to minor embed a graph onto the hardware graph of a D-Wave annealer~\cite{Yarkoni2021}.

As well as the minor embedding cost, another drawback of the quadratic penalty method is that it will often reduce the effective dynamic range of qubit interactions. Taking the quadratic penalty for the constraints~\ref{constraint:C1} and expanding out the brackets, we get
\begin{eqnarray}
    \nonumber
    P(\mathbf{x}) &=& \alpha_2 \left(\sum_{i=1}^{n_p} x_{i, q} - A \right)^2 \\
    \label{eq:C1_quadratic_penalty_function}
    &=& \alpha_2 \left( \sum_{i=1}^{n_p}(1-2A)x_{i,q} + \sum_{i=1}^{n_p-1} \sum_{j=i+1}^{n_p} 2 x_{i,q} x_{j,q} + A^2 \right).
\end{eqnarray}
Applying the QUBO to Ising mapping $x_{i,q} \mapsto (1 - \sigma_{i,q}^z)/2$ gives
\begin{eqnarray}
    \nonumber
    P &=& \alpha_2 \left( \sum_{i=1}^{n_p-1} \sum_{j=i+1}^{n_p} \frac{\sigma_{i,q}^z \sigma_{j,q}^z}{2} + \sum_{i=1}^{n_p} \left( \frac{n_p}{2} - A \right) \sigma_{i,q}^z \right. \\
    \label{eq:C1_quadratic_penalty_function_ising}
    && \left. + \frac{n_p(n_p + 1)}{4} - n_p A + A^2 \right).
\end{eqnarray}
The couplings and local fields can be read off as $J_{i,j} = \alpha_2 / 2$ and $h_i = \alpha_2 (n_p/2 - A)$. The magnitudes of these coefficients increase with the magnitude of $\alpha_2$, while the magnitude of the local fields is also proportional to the absolute difference $|n_p/2 - A|$. Therefore, if the desired constraint value (number of promotions $A$) is far from half of the number of variables in the constraint (number of products $n_p$), this penalty introduces strong local fields.

There are physical limitations on the range of $\mathbf{J}$ or $\mathbf{h}$ values that can be implemented on a quantum annealer. Because of this, the Ising Hamiltonian $\widetilde{H}_P$ implemented by the device is normalised by a factor $\mathcal{N}$ using
\begin{equation}
    \widetilde{H}_P = \frac{1}{\mathcal{N}} H_P,
    \label{eq:normalised_problem_hamiltonian}
\end{equation}
where $\mathcal{N}$ is usually chosen to be the minimum value such that all physical constraints on $\mathbf{J}$ and $\mathbf{h}$ are satisfied. A penalty that introduces large-magnitude couplings or local fields is undesirable as it will often result in a larger factor $\mathcal{N}$, reducing the effective dynamic range of qubit interactions for the unconstrained part of the problem.

Other methods of encoding constraints in QA have been proposed as alternatives to the quadratic penalty method. One approach is to engineer interactions in the driver Hamiltonian $H_D$ that only allow for transitions between feasible states, thereby naturally limiting the search space to the feasible subspace~\cite{Hen2016, Hen2016a}. This method can be combined with an encoding scheme in which qubits represent the parities of products of spin variables~\cite{Lechner2015, Drieb-Schon2023}. While the approach of re-engineering $H_D$ would avoid the mentioned limitations of the quadratic penalty method, it requires multi-qubit interactions that can be challenging to physically implement. Currently, there are no commercial quantum annealers that support these interactions. A different method uses small Ising problems called \textit{gadgets} that are designed to have properties that allow them to be combined in such a way to encode the original constrained problem~\cite{Vyskocil2019, Vyskocil2019a, Djidjev2020}. To create the gadgets, ancillary qubits are used. The layout of the gadgets can be tailored to a device's hardware, leading to a better efficiency than the quadratic penalty method in terms of dynamic range and the number of physical qubits required after minor embedding.

In this work, we consider the use of linear Ising penalty functions for linear equality constraints. \textit{Non-Ising} linear penalty functions have been used by classical solvers~\cite{Fletcher1983}. For example, the penalty function $\alpha_1 \left| \sum_{i=1}^n \mu_i x_i - c \right|$ implements the general linear equality constraint in Eq.~\ref{eq:linear_equality_constraint}.  Computing this type of penalty on a quantum computer requires the introduction of ancillary qubits because of the non-Ising operation $|\cdot|$. This has been demonstrated in the context of the quantum approximate optimization algorithm~\cite{DelaGrandrive2019}. Instead, we remove the non-Ising operator and use a linear Ising penalty function of the form
\begin{equation}
    P(\mathbf{x}) = \alpha_1 \left( \sum_{i=1}^{n_p} x_{i,q} - A \right)
    \label{eq:C1_linear_penalty_function}
\end{equation}
for the equality constraints in Eq.~\ref{eq:constraint_C1}. Here, $\alpha_1$ is a penalty strength that can be positive or negative. Due to the removal of the non-Ising operation, property~\ref{item:penalty_function_property_2} of a conventional penalty function is no longer fulfilled. Therefore, this penalty method is not always successful in producing a feasible solution as the ground state of $H_P$. We make this trade-off so that the penalty function can be implemented with no ancillary qubits or couplings, thereby avoiding many of the drawbacks of the quadratic penalty method. The linear Ising penalty method has previously been suggested for QA in the contexts of portfolio optimization~\cite{Venturelli2019} and quantum machine learning~\cite{Willsch2020a}. In the rest of this paper, the term \textit{linear penalty} is used to refer to linear Ising penalties of the form in Eq.~\ref{eq:C1_linear_penalty_function} rather than linear penalties with non-Ising operations.

Recent work by Ohzeki~\cite{Ohzeki2020} demonstrates a method for implementing constraints with linear terms by applying a Hubbard-Stratonovich transformation~\cite{stratonovich1957method, Hubbard1959} to the partition function of a QUBO objective function. Although Ohzeki uses a different mathematical motivation to what we present here, their work arrives at an algorithm that is effectively the same as applying linear penalties, where $\alpha_1$ in Eq.~\ref{eq:C1_linear_penalty_function} is viewed as a Lagrange multiplier. This method is sometimes unsuccessful in exactly implementing hard constraints~\cite{Kuramata2021}. In this work, we quantify the prevalence of instances that cannot be exactly constrained by the linear penalty method, which previous work does not do. The problems considered in this work have objective functions with non-negative quadratic coefficients, which gives them a structure that makes them particularly amenable to the linear penalty method~\cite{theorypaper}. For cases in which sampled solutions do not satisfy all constraints, a post-processing algorithm has been proposed that obtains feasible solutions from infeasible solutions while minimising the number of bit flips performed on the solution~\cite{Kuramata2021}. In this work, we use a different strategy of switching to using quadratic penalties for constraints that are not being satisfied while applying linear penalties to the other constraints.

The linear penalty function in Eq.~\ref{eq:C1_linear_penalty_function} corresponds to the application of local fields equal to $-\alpha_1/2$ in the Ising formulation, up to a constant offset. Since this does not contribute any additional couplings to $H_P$, the resulting minor embedding will often use fewer physical qubits than when the quadratic penalty method is used. Furthermore, the maximum coupling strength is unchanged, and the magnitude of the local fields that are applied by the penalty are not dependent on the values of $\alpha_2$ and $|n_p/2 - A|$, but only on $\alpha_1$. This can result in a smaller normalisation factor $\mathcal{N}$, which typically results in better performance. Such improvements in effective dynamic range and performance have been observed in numerical simulation~\cite{theorypaper}. Since in practice the linear penalties will have to be found iteratively, as discussed in detail in Ref.~\cite{theorypaper}, the use of linear penalties can be viewed as a hybrid quantum-classical algorithm~\cite{Callison2022}.

The amount by which the linear penalty in Eq.~\ref{eq:C1_linear_penalty_function} penalises a solution depends only on the Hamming weight of the variables involved in the penalty, which is the number of those variables equal to one in a solution. A linear penalty with a negative value of $\alpha_1$ energetically favours states in which the variables involved with the constraint have a larger Hamming weight, whereas a positive $\alpha_1$ favours smaller Hamming weights. To implement a constraint, $\alpha_1$ should be tuned such that the ground state of $H_P$ has the desired Hamming weight. As previously mentioned, it is not guaranteed that such a value of $\alpha_1$ exists.

It is important to consider the type of constraint and structure of the objective function when deciding whether the linear penalty method is suitable. Due to our assumption that $C$ is non-negative, the example promotion cannibalization problems in this work have objective functions comprised solely of non-negative quadratic coefficients. In Ref.~\cite{theorypaper}, it is explained that problems with this structure are more often able to be successfully constrained with the linear penalty method compared to random QUBO problems with both positive and negative couplings. Another important aspect of our examples is that the constraints we encode with linear penalties are equality constraints on the Hamming weights of variables. It is not clear whether the linear penalty method is as effective for other types of linear constraints. As a rule of thumb, Hamming weight equality constraints on variables that only have non-negative quadratic term coefficients in the objective function are good candidates for linear penalties, but this does not mean that other types of constraints aren't good candidates.

One example problem where we do not expect the linear penalty method to always be effective is the knapsack problem, which has a linear objective function and a single linear inequality constraint. This constraint differs to our example problems in that it is not a function of Hamming weight and is not an equality. If a linear penalty could implement the constraint, the penalised objective function would be trivial to solve as it would be entirely linear. Assuming the linear penalty strength could be tuned in polynomial time, this would render the NP-hard problem trivial and thus show that $\mathrm{P}=\mathrm{NP}$. Note that typical knapsack problem instances are known to be easy~\cite{Beier2004}, so it is possible that the linear penalty method works well for most, but not all, instances without requiring that $\mathrm{P}=\mathrm{NP}$. In fact, creating hard knapsack problem instances is an active area of research~\cite{Jooken2022}. Further work is required to assess how effective the linear penalty method is for other types of constraints.

The constraints~\ref{constraint:C2} and~\ref{constraint:C3} are inequality constraints, which we encode with quadratic penalty functions. For the constraints~\ref{constraint:C3}, penalties of the form
\begin{equation}
    P(\mathbf{x}) = \alpha_2 x_{i,q} x_{i,q+1}
    \label{eq:C3_quadratic_penalty_function}
\end{equation}
are sufficient. For the constraints~\ref{constraint:C2} with a range $\Delta = B_\mathrm{max}-B_\mathrm{min}$, binary encoded slack variables $\mathbf{s} \in \{0, 1\}^{4\log_2(\Delta+1)}$ can be used to encode these four constraints with penalty functions of the form
\begin{equation}
    P(\mathbf{x}, \mathbf{s}) = \alpha_2 \left( \sum_{i=1}^{n_p} x_{i,q} - B_\mathrm{max} + \sum_{j=1}^{\log_2(\Delta+1)} 2^{j-1} s_{j,q} \right) ^2
    \label{eq:C2_quadratic_penalty_function}
\end{equation}
for cases where $\Delta + 1$ is a power of two.

\section{\label{sec:numerical_and_experimental_methods}Numerical and experimental methods}

This work made extensive use of the Python programming language~\cite{VanRossum1995} along with the NumPy~\cite{harris2020array} and SciPy~\cite{2020SciPy-NMeth} libraries to perform computationally intensive calculations and Matplotlib~\cite{hunter2007matplotlib} to produce plots. The weighted least-squares method implementation in \texttt{scipy.optimize.curve\_fit} was used to obtain linear fits. PyQUBO~\cite{Zaman2022} was used as a convenient tool for the formulation of QUBO and Ising problems.

Gurobi Optimizer~\cite{gurobi} was used through the GurobiPy Python interface to find optimal solutions of problem instances as well as their minimum and maximum objective values. Although Gurobi Optimizer is an exact solver, it operates at an adjustable numerical precision, which can lead to minor differences in results depending on the software version and solver parameters being used. This work used Gurobi version 10.0.2 with a single thread and all other solver parameters set to their default values.

The $C$ matrices used in our tests on the D-Wave annealer were generated with symmetric off-diagonal elements $C_{i,j} = C_{j,i}$ selected uniformly at random from the interval $[0.1, 1)$. All main diagonal elements $C_{i,i}$ were set to $0$. These $C$ matrices were made sparse by setting some of the off-diagonal elements equal to $0$. A minimum number of nonzero elements per product was chosen, and matrix elements $C_{i,j}$ were randomly selected and set to zero if both products $i$ and $j$ had more than the minimum number of nonzero elements. This was repeated until all matrix elements had been considered. In the context of $C$ matrices, we use the term ``connectivity'' to refer to the number of nonzero cannibalization interactions with other products for a given product. This is not to be confused with hardware connectivity, which refers to the physical couplings available between physical qubits. 10,000 instances with 100 products were generated for our analysis of the single-quarter promotion cannibalization problem with the minimum connectivity for each product set to $3$. Since some products will have more than the minimum number of nonzero elements, the average connectivity is $\approx3.4$ for these instances. Another 10,000 instances were generated for the D-Wave runs on the four-quarter problem with 10 products. For these, the minimum connectivity was set to $5$ and the average connectivity is $\approx 5.1$.

QA experiments were conducted on the Advantage\_system6.3 QPU~\cite{Advantage6-3}, which is a D-Wave Advantage quantum annealer. We used the D-Wave Ocean SDK~\cite{OceanSDK} to interface with the annealer. For each problem instance, 1,000 solutions were sampled by the annealer. The \texttt{find\_embedding} function in Ocean was used to calculate minor embeddings before problems were submitted to the annealer. The \texttt{find\_embedding} function takes a seed parameter as an input, which affects the resulting embedding. We set this parameter to a different value for each problem instance for all experiments other than in the case of experiments on the single-quarter promotion cannibalisation problem using the quadratic penalty method, where the first 100 instances' minor embeddings were reused for the rest of the problem instances to save computation time. For cases where the problem was fully connected, we considered using the \texttt{find\_clique\_embedding} function instead; however, we found that although this heuristic provided more efficient minor embeddings, using \texttt{find\_embedding} resulted in better performance on average. We explain why we believe this occurs in \ref{app:clique_embedding_comparison}. Therefore, all results that we present in this paper are with the minor embeddings calculated with \texttt{find\_embedding}. Chain strengths were calculated by the \texttt{uniform\_torque\_compensation} function in D-Wave Ocean with a prefactor of 1.414, which is the default behaviour in Ocean.

To determine the value of the quadratic penalty strength $\alpha_2$ used in experiments, the simulated annealing algorithm~\cite{Kirkpatrick1983} was utilised. Simulated annealing runs were performed on the Hamilton high performance computing cluster at Durham University with the \texttt{SimulatedAnnealingSampler} sampler in Ocean. The solver was set to sample 1,000 solutions with a random seed of 0. All other solver parameters were set to their default values.

\section{\label{sec:minor_embedding_improvements}Minor embedding improvements}

\begin{figure*}
    \centering
    \includegraphics[width=\textwidth]{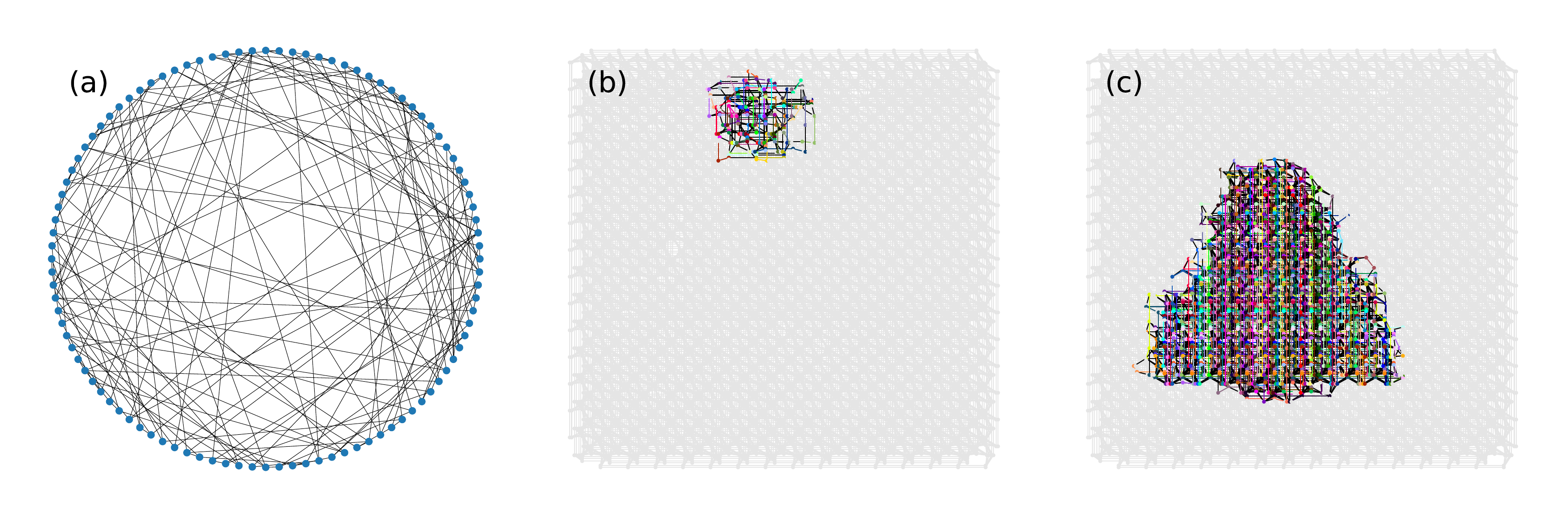}
    \caption{(a) Logical graph of an instance of the single-quarter promotion cannibalization problem with $n_p = 100$ products and without any penalty functions applied. Nodes represent variables and edges represent couplings. (b) An example of a minor embedding of the problem instance onto the working graph of the D-Wave advantage QPU after applying a linear penalty. The linear penalty does not change the logical graph of the problem. Nodes representing physical qubits are coloured based on the logical variable they represent. Black edges represent couplings between different variables. Qubits and couplings that are unused are coloured in grey. (c) An example of a minor embedding of the same problem instance where the quadratic penalty method is used instead. The quadratic penalty results in a fully connected logical graph, which requires more physical qubits to minor embed than with a linear penalty.
    }
    \label{fig:minor_embedding_examples}
\end{figure*}

As mentioned in Sec.~\ref{sec:background}, one of the advantages of using the linear penalty method over the quadratic penalty method is that it does not introduce any new couplings that don't already exist in $H_P$. For QA on D-Wave hardware, this results in fewer physical qubits being used after minor embedding. In order to demonstrate this difference in minor embedding efficiency between the linear and quadratic penalty methods when solving the single-quarter problem, we plot two example minor embeddings in Fig.~\ref{fig:minor_embedding_examples}. These correspond to the logical graph of the instance with ID \texttt{100\_0} and were both calculated by the \texttt{find\_embedding} function in D-Wave Ocean. The minor embedding with the linear penalty method uses 189 physical qubits, whereas $1{,}282$ physical qubits are used with the quadratic penalty method. This substantial difference arises because the quadratic penalty requires a fully connected logical graph, whereas the linear penalty maintains the sparsity of the logical graph.

\begin{figure}
    \centering
    \includegraphics[width=0.6\columnwidth]{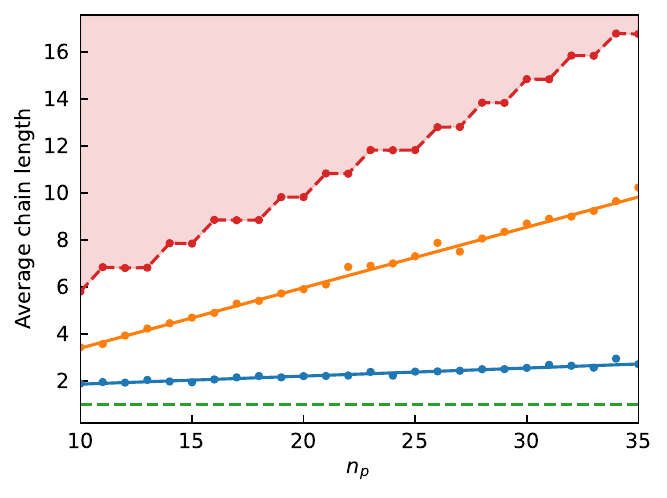}
    \caption{
    Average chain length after minor embedding a four-quarter promotion cannibalization problem onto the D-Wave Advantage hardware graph against the number of products $n_p$ in the problem. We consider the case of an inequality constraint on the number of promotions per product that introduces one slack variable per product. Hence, the number of logical variables is $5 n_p$. The $C$ matrices are assumed to be sparse, with a minimum connectivity of 3. We compare the use of quadratic penalties (orange) against linear penalties (blue) for the constraints on the number of promotions per quarter, using the \texttt{find\_embedding} function to calculate the minor embeddings. These points are averaged over 10 possible structures of the $C$ matrix and given linear fits. We also plot the average chain length after embedding with the \texttt{find\_clique\_embedding} function in red, with the red shaded region representing the regime in which switching to this embedding heuristic would produce an embedding that uses fewer physical qubits. The green line represents the chain length if the problem could be directly mapped to the hardware (i.e.\ 1).
    }
    \label{fig:minor_embedding_scaling}
\end{figure}

To quantify how the minor embedding efficiency of the penalty methods scale with problem size, we have calculated minor embeddings of the four-quarter promotion cannibalization problem onto the hardware graph of the D-Wave Advantage\_system6.3 for various different numbers of products. In Fig.~\ref{fig:minor_embedding_scaling}, we show how the average chain length (i.e.~the number of physical qubits per logical variable) after minor embedding scales with the problem size for problems with sparse $C$ matrices. We observe that the average chain length scales much more favourably when using linear penalties for the constraints on the number of promotions per quarter than when using quadratic penalties.

As the number of physical qubits of a quantum annealer is limited, it is not always possible to find a minor embedding that fits onto the hardware. By producing minor embeddings with fewer physical qubits, the linear penalty method allows for larger problems to be tackled than what is possible with the quadratic penalty method. When using the annealer to solve multiple problems in parallel, smaller embeddings make it possible to embed more problem instances on the hardware, which saves compute time. We have not attempted this type of parallelisation in our experiments.

The reduction in average chain length when using the linear penalty method may also lead to better performance in minimising the objective function. As the chain length for a particular logical variable is increased, it becomes less likely that all associated physical qubits are measured to be in the same state at the end of an anneal. In cases where there is a disagreement between the physical qubits, which is called a \textit{chain break}, some strategy is applied to interpret the value of the logical qubit. A common strategy is a majority vote. Shorter chains have been observed to produce better performance~\cite{King2020, Haghighi2023}.

\section{\label{sec:dwave_results}Performance on a D-Wave quantum annealer}

We have evaluated the performance of the D-Wave Advantage\_system6.3 quantum annealer in solving promotion cannibalization problems using the linear and quadratic penalty methods, and we present our results in this section. In Sec.~\ref{sec:single_quarter_dwave_results}, we consider the single-quarter promotion cannibalization problem, and in Sec.~\ref{sec:four_quarter_dwave_results}, we consider the four-quarter problem.

\subsection{\label{sec:single_quarter_dwave_results}Problem with a single constraint}

In this subsection, we consider the formulation of the promotion cannibalization problem in which we want to find a promotion plan for a single fiscal quarter. We examine the case where there are $n_p = 100$ products to choose from. The only constraint in this problem is the desired number of promotions $A$, which we choose to be 50. While it is unusual for a retailer to promote half of their available products in a single quarter, we can imagine a more realistic scenario in which the 100 products in the problem are a selection of products that have been determined to be the most promising products to promote out of a larger set of available products. In this scenario, the quantum optimizer is being used to select from these 100 candidate promotions. The $C$ matrices we have used for this problem have an average connectivity of $\approx3.4$. Even at this level of sparsity, the optimal solutions of all the constrained problem instances we generated have nonzero total cannibalization.

\begin{figure}
    \centering
    \includegraphics[width=\columnwidth]{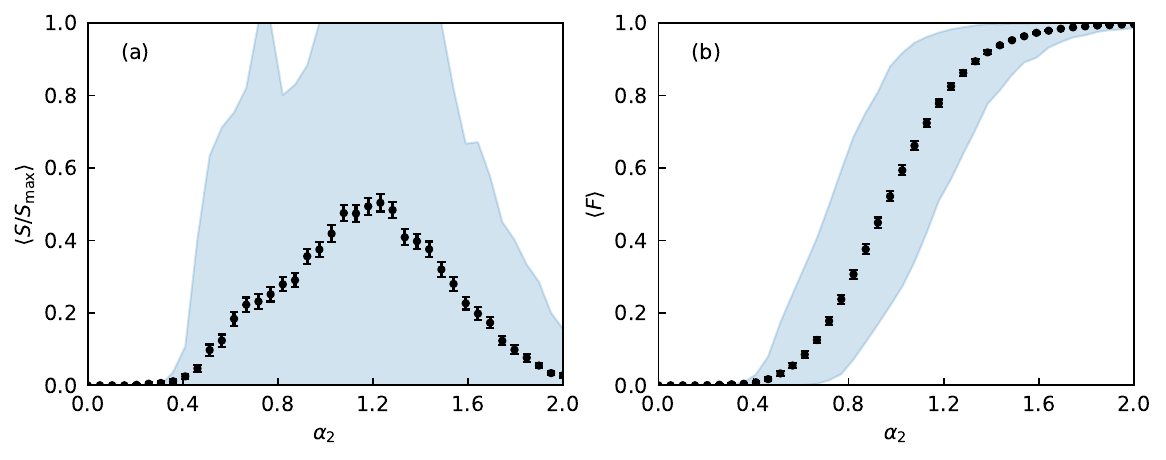}
    \caption{(a) For a simulated annealing algorithm solving the single-quarter promotion cannibalization problem with different values of the penalty parameter $\alpha_2$, we plot the fraction $S$ of sampled solutions that are optimal normalised by the maximum measured fraction $S_\mathrm{max}$. The points are averaged over 200 instances of the problem, with error bars representing the standard error in the mean. The blue shaded region contains the 5th to 95th percentile values of $S/S_\mathrm{max}$. Note that $S_\mathrm{max}$ is calculated for each instance separately and is a maximum over the plotted values of $\alpha_2$. (b) Same as (a), with the y axis instead showing the fraction $F$ of sampled solutions that are feasible.}
    \label{fig:1x100_quadratic_penalty_strength}
\end{figure}

The quadratic and linear penalty methods behave differently when their penalty strengths $\alpha_2$ and $\alpha_1$ are changed. For the quadratic method, $\alpha_2$ must be large enough to produce a feasible ground state in $H_P$ but not so large that it hinders the quantum dynamics. Typically, there is a wide range of $\alpha_2$ values that produce near-optimal performance. Fig.~\ref{fig:1x100_quadratic_penalty_strength} and Fig.~\ref{fig:4x10_quadratic_penalty_strengths} show examples of this behaviour in the case of simulated annealing. In comparison, for the linear penalty method, the value of $\alpha_1$ determines the value of the constraint. More precisely, the value of $A$ in Eq.~\ref{eq:constraint_C1_single_quarter} increases monotonically as $\alpha_1$ is decreased. Therefore, performance can often be more sensitive to the value of $\alpha_1$. This is explored in more detail in Ref.~\cite{theorypaper}.

As suggested in Ref.~\cite{Venturelli2019}, a value of $\alpha_1$ that correctly implements the constraint in the single-quarter problem can be found with a simple search strategy that samples solutions and decreases (increases) $\alpha_1$ if the Hamming weight of the best found solution is too low (high). This strategy relies on the monotonic relationship between $\alpha_1$ and $A$. For some problem instances, there is no value of $\alpha_1$ that produces the desired constraint. Ref.~\cite{theorypaper} provides a more detailed discussion of searching for $\alpha_1$ in problems with a single linear penalty and why a good value cannot be found for some instances\footnote{Ref.~\cite{theorypaper} shows that by only having non-negative coefficients in the objective function of this promotion cannibalization problem, the minimum objective value of solutions monotonically increases with Hamming weight. This structure makes it more likely that a constraint can be exactly implemented with a linear penalty, but the gradient of a plot of minimum objective value against Hamming weight would also need to be monotonically increasing to \textit{guarantee} that a constraint can always be implemented, which isn't always the case for this problem.}. In our analysis, we choose a value of $\alpha_1$ for each instance uniformly at random from the range of values that produce the correct ground state (if it exists and could be found up to a finite precision). These are not necessarily the optimal values of $\alpha_1$; instead, they approximately represent the typical $\alpha_1$ values that would be found by a search strategy. Out of 10,000 problem instances that were randomly generated, there were 1,406 instances for which a value of $\alpha_1$ that produced the correct ground state could not be found. The D-Wave runs in this subsection were performed on 1,000 instances that were randomly selected from the remaining 8,594 instances for which the linear penalty method was able to produce the desired constraint.

Since the quadratic penalty method is not as sensitive to the value of the penalty strength as the linear penalty method, using a single value of $\alpha_2$ for all instances often works well in practice. For all single-quarter problem instance considered here, we set $\alpha_2=1.2$. We arrived at this value by considering the fraction of solutions sampled by a simulated annealing algorithm that are optimal or feasible for 200 randomly selected problem instances as a function of $\alpha_2$, which is plotted in Fig.~\ref{fig:1x100_quadratic_penalty_strength}. The fraction $S$ of optimal solutions was normalised by the maximum value $S_\mathrm{max}$ for each instance to prevent the average from being dominated by a few instances for which $S$ is large. For the sake of transparency, we show the data without normalisation in \ref{app:1x100_quadratic_penalty_strength_choice}. In Fig.~\ref{fig:1x100_quadratic_penalty_strength}(a), we can see that the average value of $S/S_\mathrm{max}$ peaks near $\alpha_2=1.2$. Fig.~\ref{fig:1x100_quadratic_penalty_strength}(b) shows that $\alpha_2=1.2$ also produces a large fraction $F$ of solutions that are feasible for the majority of instances. Although the large confidence interval in Fig.~\ref{fig:1x100_quadratic_penalty_strength}(a) indicates that there are some instances for which $\alpha_2=1.2$ does not produce a near-optimal probability of sampling an optimal solution, the fact that the average value of $S/S_\mathrm{max}$ is around $0.5$ implies that $S$ is not far from its maximum value for the majority of instances. While we do not expect the D-Wave quantum annealer to behave exactly the same as the simulated annealing algorithm considered here, the broadness of the peaks in Fig.~\ref{fig:1x100_quadratic_penalty_strength}(a) and Fig.~\ref{fig:4x10_quadratic_penalty_strengths}(a-c) suggests that our results from the quantum annealer would not be substantially impacted by small changes in our choice of $\alpha_2$.

In this example problem, a quadratic penalty introduces many new couplings, and this has a large impact on the minor embedding, as seen in the example in Fig.~\ref{fig:minor_embedding_examples}. Since the quadratic penalty acts on all variables, it produces a logical graph that is fully connected. For the 1,000 problem instances that we have used in our D-Wave tests, $\approx 1{,}286$ physical qubits are used on average to minor embed the 100 logical variables when the quadratic penalty method is used. In contrast, no new couplings are introduced when the linear penalty method is used, so the only couplings that exist come from the matrices $C$, which are sparse. After minor embedding, $\approx 156$ physical qubits are used on average when the linear penalty method is used for the same problem instances. This is a very significant reduction in the number of physical qubits, and we expect it to improve performance.

Since the constraint value of $A=50$ promotions is exactly half of the total number of products $n_p$ in this problem, the quadratic penalty for this constraint (Eq.~\ref{eq:C1_quadratic_penalty_function_ising}) applies no local fields and only contributes to the couplings. This means that the dynamic range improvements from switching to linear penalties are not as significant as they would be for other constraint values, where the quadratic penalty would introduce strong local fields. For the problem instances used in this subsection, the maximum coupling strength $\max(|J_{i,j}|)$ is $\approx 2.21 \times$ larger and the maximum local field strength $\max(|h_i|)$ is $\approx 1.26 \times$ larger when the quadratic penalty method is used than when the linear method is used. Therefore, switching to the linear penalty method is expected to improve the dynamic range of qubit interactions by reducing the value of $\mathcal{N}$ in Eq.~\ref{eq:normalised_problem_hamiltonian}. We suspect that this improvement does not impact performance as significantly as the improvement in minor embedding efficiency in this experiment.

\begin{figure}
    \centering
    \includegraphics[width=\columnwidth]{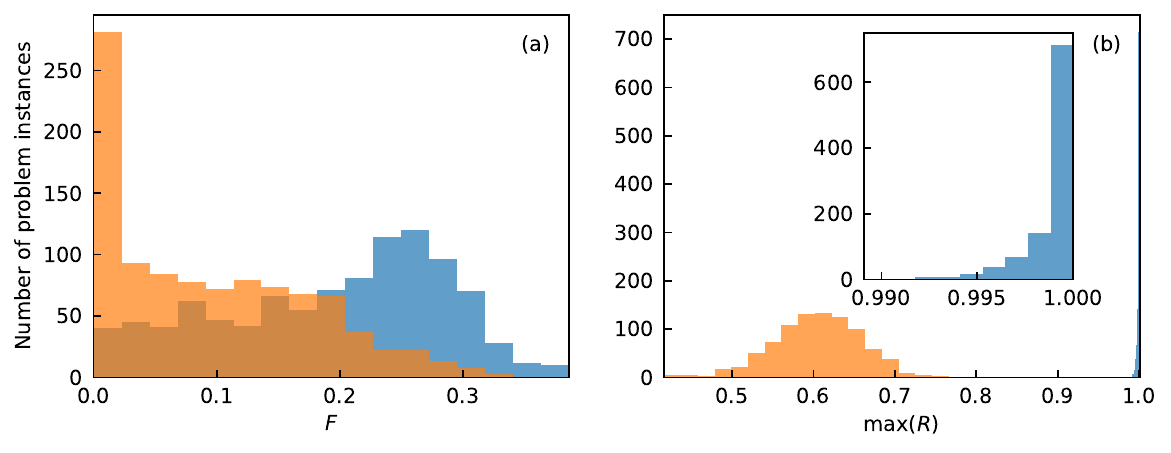}
    \caption{(a) Histograms of the fraction $F$ of 1,000 D-Wave annealer samples that are feasible solutions for 1,000 instances of the single-quarter promotion cannibalization problem. Samples were taken using the linear penalty method (blue) and the quadratic penalty method (orange) to encode the problem's constraint.
    (b) Histograms of the maximum approximation ratios $R$ of the same D-Wave sample sets. Note that the two histograms are using different bin widths to make the shapes of their distributions clearer. The inset zooms into the region showing the histogram for the linear penalty method so it can be seen more clearly.}
    \label{fig:1x100_results}
\end{figure}

The results of the D-Wave annealer runs on the single-quarter problem are shown in Fig.~\ref{fig:1x100_results}. In Fig.~\ref{fig:1x100_results}(a), we can see a noticeable increase in the fraction of sampled solutions that are feasible when the quadratic penalty is replaced with a linear penalty. Out of the 1,000 problem instances, there were 117 instances that the quantum annealer did not sample any feasible solutions for using the quadratic penalty method, compared to only 4 instances with the linear penalty method. Fig.~\ref{fig:1x100_results}(b) shows histograms of the approximation ratios $R$ of the best feasible solutions that were sampled. With the quadratic penalty method, the quantum annealer is only able to sample low-quality feasible solutions, whereas with the linear penalty method, the annealer finds optimal and near-optimal solutions. Optimal solutions were found for 541 of the instances using the linear penalty method. There is a clear advantage of using the linear penalty method over the quadratic method in this example, which we presume is primarily due to the improvements in minor embedding efficiency.

\subsection{\label{sec:four_quarter_dwave_results}Problem with multiple constraints}

In this subsection, we consider the promotion cannibalization problem with four fiscal quarters, $n_p = 10$ products, and multiple constraints. The seasonal scale factors in Eq.~\ref{eq:four_quarter_promotion_cannibalization_qubo_objective_function} are set to $\boldsymbol{\lambda}=(1.5, 1.0, 1.0, 1.5)^\mathrm{T}$ for all problem instances. We use either the linear or quadratic penalty method for the constraints~\ref{constraint:C1} on the number of promotions per quarter, while constraints~\ref{constraint:C2} and~\ref{constraint:C3} are implemented with the quadratic penalty method only. For the constraints \ref{constraint:C1}, we have set $A=4$, which requires four out of the ten products to be promoted in each quarter. For the constraints \ref{constraint:C2}, we have set $B_\mathrm{min} = 1$ and $B_\mathrm{min} = 2$, which requires each product to be promoted once or twice in total. Note that this introduces $10$ extra variables with the slack variable encoding used in Eq.~\ref{eq:C2_quadratic_penalty_function}, bringing the total variable count to $50$ for this problem.

The $C$ matrices used in this problem have an average connectivity of $\approx 5.1$, making them more dense than those used in the single-quarter problem. This is due to the smaller number of promotions per quarter in this problem, which results in optimal solutions with zero cannibalization occurring frequently if the $C$ matrices are made more sparse. Our choice of problem size is limited by the performance of the D-Wave annealer. In real-world problems, the number of products and promotions would be much larger, so sparser $C$ matrices would still produce optimal solutions with non-zero cannibalization.

In contrast to the single-quarter promotion cannibalization problem, the four-quarter problem has many constraints that can each take a different value of the penalty strength. In our analysis, we used a different value of $\alpha_2$ for each of these three sets of constraints \ref{constraint:C1}, \ref{constraint:C2}, and \ref{constraint:C3}. In line with the previous subsection, the $\alpha_2$ values were chosen using simulated annealing runs. The results of these runs and our chosen values of $\alpha_2$ are shown in \ref{app:4x10_quadratic_penalty_strength_choices}.

When there are multiple constraints that we are using the linear penalty method for, the different linear penalties can interact with each other, making the search for $\alpha_1$ values more difficult. In other words, changing the penalty strength for a particular constraint does not just change the assignment of the variables within the constraint, but it can also change the assignment of other variables through their couplings. Nevertheless, we have implemented a simple search strategy that, when paired with a solver that returns the optimal solution, works well in practice for the four-quarter problem instances. This search strategy is detailed in \ref{app:linear_penalty_search_strategy}. We are not certain whether a search strategy that can efficiently find penalty strengths that produces a feasible ground state (up to a desired precision) exists for the four-quarter problem. However, we suspect that the fact that the problem has non-negative couplings makes it likely that such a strategy exists, as in the case for the single-quarter problem.

We first compare the use of linear penalties for all four constraints~\ref{constraint:C1} against the use of all-quadratic penalties. Out of the 10,000 randomly generated problem instances, we were able to find values of $\alpha_1$ that produced feasible optimal solutions for 6,066 of them. For 3,082 of these instances, using the same value of $\alpha_1$ for each of the four linear penalties was successful. In this set of D-Wave runs, we used 1,500 instances that were randomly selected from the 3,082 instances that can be constrained with the same $\alpha_1$ parameter for each quarter. The specific value of $\alpha_1$ for each instance was chosen uniformly at random from the range of values that produce feasible ground states, like in the previous subsection.

\begin{figure}
    \centering
    \includegraphics[width=\columnwidth]{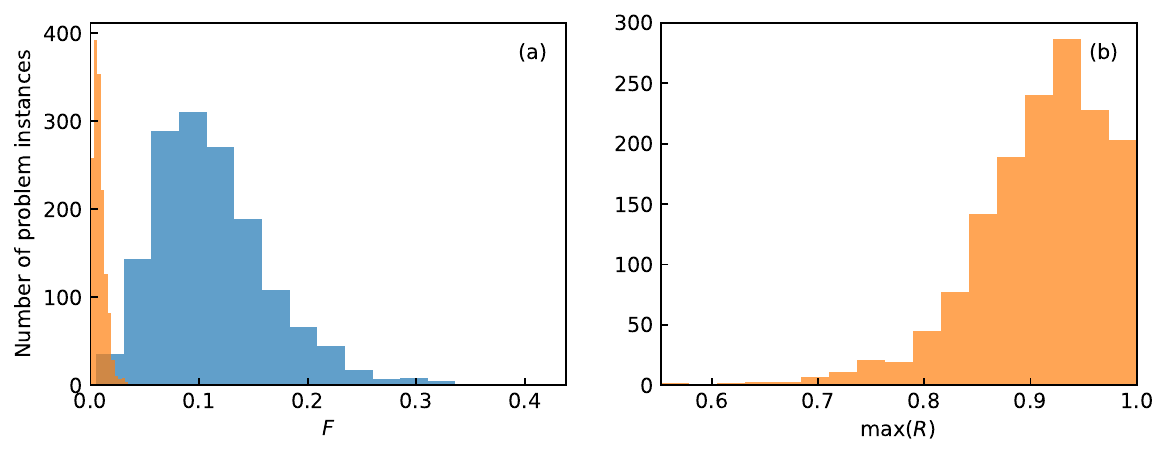}
    \caption{(a) Histograms of the fraction $F$ of 1,000 D-Wave annealer samples that are feasible solutions for 1,500 instances of the four-quarter promotion cannibalization problem. Samples were taken from the D-Wave annealer using the linear penalty method (blue) and quadratic penalty method (orange) to encode the constraints on the number of promotions per fiscal quarter. Note that the two histograms are using different bin widths so that the shapes of their distributions are clearer.
    (b) Histogram of the maximum approximation ratios $R$ of the D-Wave sample sets using only quadratic penalties. A histogram for the linear penalty method sample sets is not shown because $\max(R)$ was equal to 1 for all but nine of the instances.
    }
    \label{fig:4x10_probabilities}
\end{figure}

In Fig.~\ref{fig:4x10_probabilities}, we show histograms of the fraction of samples that are feasible and the approximation ratio of the best feasible sample for both penalty methods. The quality of the solutions is significantly enhanced when switching the penalties on each quarter from quadratic to linear, which is in alignment with our previous findings for the single-quarter problem. The penalty scheme involving four linear penalties found feasible solutions for all 1,500 instances and found optimal solutions for 1,491 of them. In comparison, the scheme using all-quadratic penalties was not able to find feasible solutions for 21 of the instances and found the optimal solution for only 55 instances.

As mentioned above, there were 3,934 instances for which we could not find values of $\alpha_1$ that correctly implemented all of the constraints in the ground state of $H_P$. This is unavoidable for some problem instances, and it may happen more frequently as the problem size is increased. It will also be more common for some problems that are formulated differently to the promotion cannibalization problem. We propose that in these cases, linear penalties that aren't working as intended can be switched to quadratic penalties while keeping as many linear penalties as possible. This can be done by switching the misbehaving penalties from linear to quadratic one by one until the best sampled solution satisfies all of the constraints. To investigate whether such a method would work well in practice, we have taken the 3,934 instances for which we could not implement all four constraints~\ref{constraint:C1} with linear penalties and attempted to find linear penalty strengths for a strategy where two quarters have linear penalties applied and two have quadratic penalties applied. 2,298 of these instances were amenable to this approach.

We performed another set of  D-Wave annealer experiments on the 2,298 instances for which the linear penalty method applied to all quarters does not produce the correct ground state but a combination of linear and quadratic penalties does. Due to our choice of seasonal scale factors $\boldsymbol{\lambda}=(1.5, 1.0, 1.0, 1.5)^\mathrm{T}$, there is a symmetry in the problem that results in the objective value of a solution remaining the same if the promotion plans for the first \& fourth quarter and the second \& third quarters are swapped. Therefore, to fix linear penalties that produce the wrong Hamming weight, penalties should be switched from linear to quadratic in pairs. Specifically, the only two linear-quadratic penalty combinations that can potentially fix misbehaving penalties are those where the first and fourth quarters have quadratic penalties (QLLQ) or where the second and third quarters do (LQQL). By the same symmetry argument as before, in the cases were LQQL or QLLQ is successful in producing a feasible ground state, the two $\alpha_1$ values can always be chosen to be the same. In our D-Wave annealer runs, we used the same $\alpha_1$ value for each pair of linear penalties. As before, these values were randomly chosen from the range of $\alpha_1$ values that produce feasible ground states for each instance.

Our tests on the quantum annealer show that using quadratic penalties for all four quarters produces feasible solutions more often than when two of the penalties are quadratic and two are linear, as shown in Table~\ref{tab:feasible_fraction_4x10_partial}. The LQQL and QLLQ schemes were only applied to the subset of instances for which the corresponding scheme produced the correct ground state, which means their results are not directly comparable as the instances in each set are not the same. With this caveat in mind, we note that the LQQL scheme produced feasible solutions for a larger fraction of instances than the QLLQ scheme.

\begin{table}
    \centering
    \begin{tabular}{l@{\hspace{5mm}}c}
        \hline
        \noalign{\smallskip}
        \multirow{2}{*}{Penalty scheme} & Fraction of problem instances for which \\
        & a feasible solution was sampled \\
        \noalign{\smallskip}
        \hline
        \noalign{\smallskip}
        All-quadratic & $\approx 0.96$ \\
        LQQL & $\approx 0.75$ \\
        QLLQ & $0.60$ \\
        \noalign{\smallskip}
        \hline
    \end{tabular}
    \caption{Fraction of problem instances for which a feasible solution was found in at least one of the 1,000 D-Wave samples for the four-quarter promotion cannibalization problem. We compare the penalty scheme of applying quadratic penalties to all four quarters against the schemes of using linear penalties for the first and last quarters (LQQL) or the second and third quarters (QLLQ).}
    \label{tab:feasible_fraction_4x10_partial}
\end{table}

The objective values of the best solutions sampled with the LQQL and QLLQ schemes are very close to those when using the all-quadratic penalty method. This is unlike the results of our previous tests where the linear penalty method produced a significant improvement over the all-quadratic method. Therefore, a more rigorous analysis is required here to accurately compare the subtle differences in performance between the penalty schemes. To do this, we compare the solution objective values on a per-instance basis in order to avoid comparing results for instances that are more difficult to solve against results for easier instances. We leave out instances for which one of the penalty methods being compared failed to find any feasible solutions.

\begin{figure}
    \centering
    \includegraphics[width=\columnwidth]{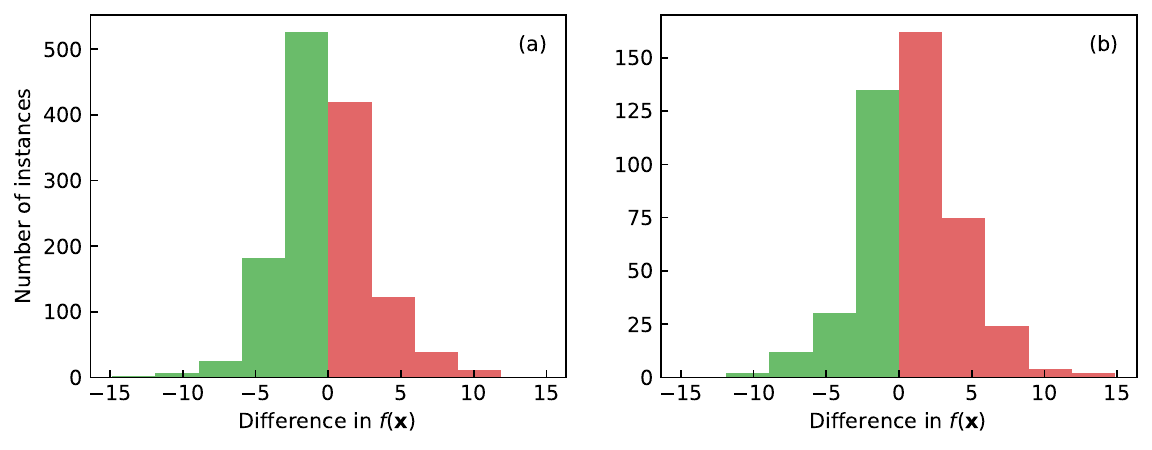}
    \caption{Difference in objective value $f(\mathbf{x})$ of the best feasible sample using the all-quadratic penalty scheme versus the (a) LQQL penalty scheme or (b) versus the QLLQ penalty scheme. A negative difference (green) indicates that the LQQL/QLLQ scheme found a higher quality solution, and a positive difference (red) indicates that the all-quadratic scheme found a higher quality solution. One instance for which the difference in best objective value was zero is left out in (a).} \label{fig:4x10_partial_energy_differences}
\end{figure}

In Fig.~\ref{fig:4x10_partial_energy_differences}, we plot histograms of the difference in objective values between the best feasible solutions sampled using the all-quadratic penalty scheme and the LQQL or QLLQ scheme. For each instance, a negative difference in $f(\mathbf{x})$ indicates that the introduction of the linear penalties resulted in a higher quality solution being sampled, and a positive difference indicates that it resulted in a lower quality solution. There were 741 instances for which the LQQL scheme improved the quality of the best solution compared to the all-quadratic scheme and 592 instances for which the LQQL scheme worsened the quality of the best solution. For the QLLQ scheme, there were 179 instances for which the scheme improved the quality of the best solution and 265 instances for which the scheme worsened the quality of the best solution.

It appears that LQQL had a slight positive impact on the solution quality whereas QLLQ had a slight negative impact, but it is unclear from this analysis alone whether these results are statistically significant. To determine this, we can perform hypothesis tests on the number of times $n_b$ that the LQQL or QLLQ scheme found a lower energy feasible solution (i.e.\ performed better) than the all-quadratic penalty scheme and the number of times $n_w$ where it could only find higher energy feasible solutions (i.e.\ performed worse). We ignore one case where the performance was the same. Our null hypothesis is that there is an equal probability of better or worse performance than the all-quadratic scheme. The statistical significance of this is given by
\begin{equation}
    p = \frac{1}{2^{n_b+n_w}} \sum_{k=n_b}^{n_b+n_w} \left( \begin{array}{c}
        n_b + n_w \\
        k
    \end{array} \right),
\end{equation}
where $\left( \begin{array}{c} n \\ k \end{array} \right)$ is the binomial coefficient. This is effectively the probability that the penalty scheme being considered would happen to perform better than the all-quadratic scheme at least as many as times if there was a 50\% chance of it performing better for any given instance. Conventionally, $p<0.05$ indicates a statistically significant result that rejects the null hypothesis~\cite{Lehmann1993} and confirms that the penalty scheme under consideration performs better than the all-quadratic scheme. We can also use this as a test for whether the penalty scheme performs worse than the all-quadratic scheme. We define $\widetilde{p}=1-p$. A value of $\widetilde{p}<0.05$ is also a statistically significant result, and it confirms worse performance than the all-quadratic scheme.

Substituting $n_b=741$ and $n_w=592$ gives $p = 2.5\times10^{-5}$, which confirms the statistical significance of our result that the LQQL scheme finds higher quality solutions for more instances than the all-quadratic scheme. Substituting $n_b=179$ and $n_w=267$ gives $\widetilde{p} = 1.2\times10^{-5}$, which shows that the result indicating that the QLLQ scheme finds higher quality solutions for fewer instances than the all-quadratic scheme is also statistically significant.

Why does switching half of the constraints from quadratic to linear penalties have such a small effect on performance compared to switching all penalties to linear? We suspect that the reason for this is that the dynamic range improvements from using linear penalties are a significant factor in improving performance in this example problem, and having a single quadratic penalty on one of the quarters is enough to eliminate most of these dynamic range improvements. Each Hamiltonian is normalised based on its largest magnitude values of $J_{i,j}$ and $h_i$. In some scenarios, a single quadratic penalty is enough to bring the largest magnitude $J_{i,j}$ or $h_i$ value up by as much as if all penalties in the problem were made quadratic. In this example problem, if a quadratic penalty is placed on a quarter with the largest scale factor $\lambda_q=1.5$, the largest magnitude $J_{i,j}$ and $h_i$ values are the same as if all four quarters had quadratic penalties. If placed on one of the quarters with $\lambda_q=1$, there will be a less severe normalisation of $H_P$ than if all penalties were quadratic, but this still wouldn't be as efficient with the dynamic range as using linear penalties for all four quarters.

Indeed, we observed that $\max(|J_{i,j}|)$ and $\max(|h_i|)$ are both unchanged when using the QLLQ scheme instead of the all-quadratic scheme because the quarters with large values of $\lambda_q$ have quadratic penalties applied in both cases. When using the LQQL scheme, $\max(|J_{i,j}|)$ is reduced by $\approx 1.14 \times$ and $\max(|h_i|)$ is reduced by $\approx 1.13 \times$ on average compared to using all-quadratic penalties. This is a small improvement compared to what we observed when using linear penalties for all quarters, where $\max(|J_{i,j}|)$ was reduced by $\approx 2.66 \times$ and $\max(|h_i|)$ was reduced by $\approx 2.25 \times$ on average compared to the all-quadratic scheme. This suggests that the effective dynamic range when using the LQQL or QLLQ schemes is closer to that when using the all-quadratic scheme than when using the linear penalty method for all quarters. It supports our reasoning for why a more significant performance enhancement isn't observed with LQQL or QLLQ in this case.

For this example problem, the LQQL and QLLQ schemes are instead more successful in improving the minor embedding efficiency. Averaged over all instances that each penalty scheme was applied to, the number of physical qubits used after minor embedding is $\approx155$ for LQQL and QLLQ, compared to $\approx168$ for all-quadratic penalties and $\approx139$ for linear penalties applied to all quarters. We expect that at larger problem sizes, both the minor embedding and dynamic range improvements would be more significant and lead to a clear performance advantage in using the LQQL or QLLQ schemes. The expected improvements in dynamic range and minor embedding also depend on the formulation and sparsity of the problem, and a much larger improvement might be observed for other problems with multiple constraints.

The difference in dynamic range improvement between the LQQL and QLLQ penalty schemes is likely the reason why the LQQL scheme performed slightly better than the all-quadratic scheme in finding low-energy solutions, whereas the QLLQ scheme performed slightly worse (Fig.~\ref{fig:4x10_partial_energy_differences}). It also explains why we found that the LQQL scheme produces feasible solutions more often than the QLLQ scheme (Table~\ref{tab:feasible_fraction_4x10_partial}). Looking beyond this example promotional cannibalization problem, we infer that in order to make the most of the dynamic range improvements of the linear penalty method, the quadratic penalties that have the largest effect on dynamic range should prioritised when switching to linear penalties.

\section{\label{sec:conclusions}Conclusions}

Using a D-Wave quantum annealer, we have experimentally tested the linear penalty method for encoding equality constraints in quantum optimization. Our analysis was performed on two simplified forms of an optimization problem that is of interest in customer data science. One of the example problems contains a single constraint, whereas the other contains multiple constraints.

For most instances of the problem with a single constraint that we have considered, we have found that the linear penalty method is able to exactly implement the desired constraint, but there are some instances for which it cannot. For the problem instances with multiple constraints, applying linear penalties to four of the constraints is successful in a smaller (but still significant) fraction of instances. We found that switching some of the penalties from linear to quadratic is able to increase the fraction of instances that can be successfully constrained.

We observed a significant improvement in the performance of the quantum annealer when using the linear penalty method for the problem with a single constraint. With a quadratic penalty, the quantum annealer was not able find high quality solutions for any of the problem instances. Switching to a linear penalty function resulted in the quantum annealer finding the optimal solution for over half of the problem instances and finding near-optimal solutions for most of the other instances. Note that there are other methods of implementing constraints that we haven't considered here, and our observation of a performance improvement when using linear penalties is specifically a comparison against the quadratic penalty method.

For instances of the problem with multiple constraints in which linear penalties were successful in implementing four of the constraints, the performance of the quantum annealer was again significantly enhanced by the use of linear penalties. For problem instances that could not be successfully constrained with four linear penalties, switching two of the linear penalties to quadratic penalties resulted in a much smaller difference in performance when compared to the all-quadratic penalty scheme. We found that there is a small but statistically significant improvement in the quality of sampled solutions when the two linear penalties are applied to the constraints that have a larger impact on dynamic range, despite having a negative impact on the feasibility of solutions. This suggests that a strategy of focusing on the penalties that have the biggest impact on the dynamic range is worthwhile, which is something that can be determined without actually performing any anneals. At larger problem sizes, where the minor embedding and dynamic range benefits of linear penalties are greater, we predict a more substantial performance advantage of linear over quadratic penalties.

The linear penalty method is more sensitive to its penalty strength parameter than the quadratic penalty method, so there is often a need to make multiple calls to the solver in order to optimize the penalty strength for each problem instance. This is usually not necessary for the quadratic penalty method. However, given that quantum optimization is concerned with NP-hard problems that are expected to take exponentially long to solve, we argue that running the solver more times is often a good trade-off to make for better performance in sampling high quality solutions.

In conclusion, our findings indicate that the linear penalty method could play a role in enhancing the performance of quantum optimization algorithms. Future research investigating problems with different structures to those that we considered in this work would be beneficial in identifying the most suitable applications of the linear penalty method. As quantum computing hardware matures, it would be interesting to test the performance of the linear penalty method at larger problem sizes and on different hardware platforms.

\section*{Data availability statement}

The data and code that support the findings of this study are openly available at the following URL/DOI: \url{http://doi.org/10.15128/r2j6731386t}.

\section*{Acknowledgments}

We thank Pim van den Heuvel for providing helpful feedback. NC was supported by UK Engineering and Physical Sciences Research Council (EPSRC) grant number EP/T026715/2 (CCP-QC). PM was supported by UK EPSRC Doctoral Training Funds awarded to Durham University (EP/T518001/1) in partnership with dunnhumby. This work used D-Wave quantum annealing credits funded by BEIS capital funding via UK EPSRC capital grant EP/Y008618/1, with access managed by the Institute for Computational Cosmology at Durham University on behalf of the UK STFC DiRAC facility and the ExCALIBUR Hardware and Enabling Software scheme. DiRAC is part of the National e-Infrastructure.

\clearpage

\section*{References}
\bibliographystyle{iopart-num-mod}

\begin{thebibliography}{10}
\expandafter\ifx\csname url\endcsname\relax
  \def\url#1{{\tt #1}}\fi
\expandafter\ifx\csname urlprefix\endcsname\relax\def\urlprefix{URL }\fi
\providecommand{\eprint}[2][]{\url{#2}}

\bibitem{Preskill2018}
Preskill J 2018 {\em Quantum\/} \href{http://dx.doi.org/10.22331/q-2018-08-06-79}{{\bf 2} 79}

\bibitem{Au-Yeung2023}
Au-Yeung R, Chancellor N and Halffmann P 2023 {\em Frontiers in Quantum Science and Technology\/} \href{http://dx.doi.org/10.3389/frqst.2023.1128576}{{\bf 2} 1128576}

\bibitem{Orus2019}
Or{\'{u}}s R, Mugel S and Lizaso E 2019 {\em Reviews in Physics\/} \href{http://dx.doi.org/10.1016/j.revip.2019.100028}{{\bf 4} 100028}

\bibitem{Venturelli2019}
Venturelli D and Kondratyev A 2019 {\em Quantum Machine Intelligence\/} \href{http://dx.doi.org/10.1007/s42484-019-00001-w}{{\bf 1} 17--30}

\bibitem{Fox2021}
Fox D~M, Branson K~M and Walker R~C 2021 {\em PLOS ONE\/} \href{http://dx.doi.org/10.1371/journal.pone.0259101}{{\bf 16} 1--16}

\bibitem{Kitai2020}
Kitai K, Guo J, Ju S, Tanaka S, Tsuda K, Shiomi J and Tamura R 2020 {\em Physical Review Research\/} \href{http://dx.doi.org/10.1103/PhysRevResearch.2.013319}{{\bf 2} 013319}

\bibitem{Stollenwerk2020}
Stollenwerk T, O'Gorman B, Venturelli D, Mandra S, Rodionova O, Ng H, Sridhar B, Rieffel E~G and Biswas R 2020 {\em IEEE Transactions on Intelligent Transportation Systems\/} \href{http://dx.doi.org/10.1109/TITS.2019.2891235}{{\bf 21} 285--297}

\bibitem{Johnson2011}
Johnson M~W {\em et~al.\/} 2011 {\em Nature\/} \href{http://dx.doi.org/10.1038/nature10012}{{\bf 473} 194--198}

\bibitem{Yarkoni2021}
Yarkoni S, Raponi E, B{\"{a}}ck T and Schmitt S 2022 {\em Reports on Progress in Physics\/} \href{http://dx.doi.org/10.1088/1361-6633/ac8c54}{{\bf 85} 104001}

\bibitem{Kadowaki1998}
Kadowaki T and Nishimori H 1998 {\em Physical Review E\/} \href{http://dx.doi.org/10.1103/PhysRevE.58.5355}{{\bf 58} 5355--5363}

\bibitem{Brooke1999}
Brooke J, Bitko D, Rosenbaum T~F and Aeppli G 1999 {\em Science\/} \href{http://dx.doi.org/10.1126/science.284.5415.779}{{\bf 284} 779--781}

\bibitem{Farhi2001}
Farhi E, Goldstone J, Gutmann S, Lapan J, Lundgren A and Preda D 2001 {\em Science\/} \href{http://dx.doi.org/10.1126/science.1057726}{{\bf 292} 472--475}

\bibitem{Advantage6-3}
{D-Wave Systems} 2023 {QPU-Specific Physical Properties:~Advantage\_system6.3} {D}-Wave User Manual 09-1272A-C \urlprefix\url{https://docs.dwavesys.com/}

\bibitem{Choi2008}
Choi V 2008 {\em Quantum Information Processing\/} \href{http://dx.doi.org/10.1007/s11128-008-0082-9}{{\bf 7} 193--209}

\bibitem{Chancellor2019}
Chancellor N 2019 {\em Quantum Science and Technology\/} \href{http://dx.doi.org/10.1088/2058-9565/ab33c2}{{\bf 4} 045004}

\bibitem{Chen2021}
Chen J, Stollenwerk T and Chancellor N 2021 {\em IEEE Transactions on Quantum Engineering\/} \href{http://dx.doi.org/10.1109/TQE.2021.3094280}{{\bf 2} 1--14}

\bibitem{Berwald2023}
Berwald J, Chancellor N and Dridi R 2023 {\em Philosophical Transactions of the Royal Society A: Mathematical, Physical and Engineering Sciences\/} \href{http://dx.doi.org/10.1098/rsta.2021.0410}{{\bf 381} 20210410}

\bibitem{Ohzeki2020}
Ohzeki M 2020 {\em Scientific Reports\/} \href{http://dx.doi.org/10.1038/s41598-020-60022-5}{{\bf 10} 3126}

\bibitem{Willsch2020a}
Willsch D, Willsch M, {De Raedt} H and Michielsen K 2020 {\em Computer Physics Communications\/} \href{http://dx.doi.org/10.1016/j.cpc.2019.107006}{{\bf 248} 107006}

\bibitem{theorypaper}
Mirkarimi P, Shukla I, Hoyle D~C, Williams R and Chancellor N 2024 {Quantum optimization with linear Ising penalty functions for customer data science} arXiv preprint arXiv:2404.05467 \urlprefix\url{https://doi.org/10.48550/arXiv.2404.05467}

\bibitem{Meredith2001}
Meredith L and Maki D 2001 {\em Applied Economics\/} \href{http://dx.doi.org/10.1080/00036840010015769}{{\bf 33} 1785--1793}

\bibitem{Aguilar-Palacios2021}
Aguilar-Palacios C, Munoz-Romero S and Rojo-Alvarez J~L 2021 {\em IEEE Access\/} \href{http://dx.doi.org/10.1109/ACCESS.2021.3062222}{{\bf 9} 34078--34089}

\bibitem{nocedal1999numerical}
Nocedal J and Wright S~J 2006 {\em {Numerical Optimization}\/} Springer Series in Operations Research and Financial Engineering (Springer New York) ISBN 978-0-387-30303-1 \urlprefix\url{https://doi.org/10.1007/978-0-387-40065-5}

\bibitem{VanThoai2013}
{Van Thoai} N 2013 {Solution Methods for General Quadratic Programming Problem with Continuous and Binary Variables: Overview} {\em Advanced Computational Methods for Knowledge Engineering\/} ed Nguyen N~T, van Do T and le~Thi H~A (Heidelberg: Springer International Publishing) pp 3--17 ISBN 978-3-319-00293-4 \urlprefix\url{https://doi.org/10.1007/978-3-319-00293-4_1}

\bibitem{Hen2016}
Hen I and Spedalieri F~M 2016 {\em Physical Review Applied\/} \href{http://dx.doi.org/10.1103/PhysRevApplied.5.034007}{{\bf 5} 034007}

\bibitem{Hen2016a}
Hen I and Sarandy M~S 2016 {\em Physical Review A\/} \href{http://dx.doi.org/10.1103/PhysRevA.93.062312}{{\bf 93} 062312}

\bibitem{Lechner2015}
Lechner W, Hauke P and Zoller P 2015 {\em Science Advances\/} \href{http://dx.doi.org/10.1126/sciadv.1500838}{{\bf 1} 1--5}

\bibitem{Drieb-Schon2023}
Drieb-Sch{\"{o}}n M, Ender K, Javanmard Y and Lechner W 2023 {\em Quantum\/} \href{http://dx.doi.org/10.22331/q-2023-03-17-951}{{\bf 7} 951}

\bibitem{Vyskocil2019}
Vyskocil T and Djidjev H 2019 {\em Algorithms\/} \href{http://dx.doi.org/10.3390/A12040077}{{\bf 12} 1--24}

\bibitem{Vyskocil2019a}
Vysko{\v{c}}il T, Pakin S and Djidjev H~N 2019 {Embedding Inequality Constraints for Quantum Annealing Optimization} {\em Quantum Technology and Optimization Problems\/} ({\em Lecture Notes in Computer Science\/} vol 11413) (Cham: Springer International Publishing) pp 11--22 \urlprefix\url{https://doi.org/10.1007/978-3-030-14082-3_2}

\bibitem{Djidjev2020}
Djidjev H 2020 {Automaton-based methodology for implementing optimization constraints for quantum annealing} {\em Proceedings of the 17th ACM International Conference on Computing Frontiers\/} (New York, NY, USA: ACM) pp 118--125 \urlprefix\url{https://doi.org/10.1145/3387902.3392619}

\bibitem{Fletcher1983}
Fletcher R 1983 {\em {Penalty Functions}\/} (Berlin, Heidelberg: Springer Berlin Heidelberg) pp 87--114 ISBN 978-3-642-68874-4 \urlprefix\url{https://doi.org/10.1007/978-3-642-68874-4_5}

\bibitem{DelaGrandrive2019}
de~la Grand'rive P~D and Hullo J~F 2019 {Knapsack Problem variants of QAOA for battery revenue optimisation} arXiv preprint arXiv:1908.02210 \urlprefix\url{https://doi.org/10.48550/arXiv.1908.02210}

\bibitem{stratonovich1957method}
Stratonovich R~L 1957 {On a method of calculating quantum distribution functions} {\em Soviet Physics Doklady\/} vol~2 p 416

\bibitem{Hubbard1959}
Hubbard J 1959 {\em Physical Review Letters\/} \href{http://dx.doi.org/10.1103/PhysRevLett.3.77}{{\bf 3} 77--78}

\bibitem{Kuramata2021}
Kuramata M, Katsuki R and Nakata K 2021 {Larger Sparse Quadratic Assignment Problem Optimization Using Quantum Annealing and a Bit-Flip Heuristic Algorithm} {\em 2021 IEEE 8th International Conference on Industrial Engineering and Applications (ICIEA)\/} (IEEE) pp 556--565 ISBN 978-1-6654-2895-8 \urlprefix\url{https://doi.org/10.1109/ICIEA52957.2021.9436749}

\bibitem{Callison2022}
Callison A and Chancellor N 2022 {\em Phys. Rev. A\/} \href{http://dx.doi.org/10.1103/PhysRevA.106.010101}{{\bf 106}(1) 010101}

\bibitem{Beier2004}
Beier R and V{\"{o}}cking B 2004 {\em Journal of Computer and System Sciences\/} \href{http://dx.doi.org/10.1016/j.jcss.2004.04.004}{{\bf 69} 306--329}

\bibitem{Jooken2022}
Jooken J, Leyman P and {De Causmaecker} P 2022 {\em European Journal of Operational Research\/} \href{http://dx.doi.org/10.1016/j.ejor.2021.12.009}{{\bf 301} 841--854}

\bibitem{VanRossum1995}
van Rossum G and {Drake Jr} F~L 1995 {\em {Python tutorial}\/} (Centrum voor Wiskunde en Informatica)

\bibitem{harris2020array}
Harris C~R, Millman K~J, van~der Walt S~J, Gommers R, Virtanen P, Cournapeau D, Wieser E, Taylor J, Berg S, Smith N~J {\em et~al.\/} 2020 {\em Nature\/} \href{http://dx.doi.org/10.1038/s41586-020-2649-2}{{\bf 585} 357--362}

\bibitem{2020SciPy-NMeth}
Virtanen P, Gommers R, Oliphant T~E, Haberland M, Reddy T, Cournapeau D, Burovski E, Peterson P, Weckesser W, Bright J {\em et~al.\/} 2020 {\em Nature Methods\/} \href{http://dx.doi.org/10.1038/s41592-019-0686-2}{{\bf 17} 261--272}

\bibitem{hunter2007matplotlib}
Hunter J~D 2007 {\em Computing in Science \& Engineering\/} \href{http://dx.doi.org/10.1109/MCSE.2007.55}{{\bf 9} 90--95}

\bibitem{Zaman2022}
Zaman M, Tanahashi K and Tanaka S 2022 {\em IEEE Transactions on Computers\/} \href{http://dx.doi.org/10.1109/TC.2021.3063618}{{\bf 71} 838--850}

\bibitem{gurobi}
{Gurobi Optimization, LLC} 2023 {Gurobi Optimizer Reference Manual} \urlprefix\url{https://www.gurobi.com}

\bibitem{OceanSDK}
{D-Wave Systems} 2024 {D-Wave Ocean SDK} [Online; accessed 25-March-2024] \urlprefix\url{https://docs.ocean.dwavesys.com/en/stable/index.html}

\bibitem{Kirkpatrick1983}
Kirkpatrick S, Gelatt C~D and Vecchi M~P 1983 {\em Science\/} \href{http://dx.doi.org/10.1126/science.220.4598.671}{{\bf 220} 671--680}

\bibitem{King2020}
King A~D and Bernoudy W 2020 {Performance benefits of increased qubit connectivity in quantum annealing 3-dimensional spin glasses} arXiv preprint arXiv:2009.12479 \urlprefix\url{http://arxiv.org/abs/2009.12479}

\bibitem{Haghighi2023}
Haghighi M~K and Dimopoulos N 2023 {Minimum-Length Chain Embedding for the Phase Unwrapping Problem on D-Wave's Pegasus Graph} {\em 2023 IEEE International Conference on Quantum Computing and Engineering (QCE)\/} vol~2 (IEEE) pp 318--319 \urlprefix\url{https://doi.org/10.1109/QCE57702.2023.10261}

\bibitem{Lehmann1993}
Lehmann E~L 1993 {\em Journal of the American Statistical Association\/} \href{http://dx.doi.org/10.1080/01621459.1993.10476404}{{\bf 88} 1242--1249}

\end{thebibliography}

\providecommand{\newblock}{}

\clearpage

\appendix

\section{\label{app:clique_embedding_comparison}Comparison of minor embedding heuristics for complete graphs}

Since the constraint in the single-quarter promotion cannibalization problem acts on all variables, the corresponding quadratic penalty function produces non-zero couplings between all of the problem's variables. This makes the interaction graph of $H_P$ complete (i.e.\ fully connected). D-Wave's Ocean library has a function called \texttt{find\_clique\_embedding} that can minor embed a complete graph onto the hardware graph of a D-Wave Advantage using fewer physical qubits than the \texttt{find\_embedding} function. However, we found that the performance of the annealer was slightly better on average when using the \texttt{find\_embedding} function, even though the embeddings used more physical qubits than that of \texttt{find\_clique\_embedding}. Therefore, the results we showed in Sec.~\ref{sec:dwave_results} are all using \texttt{find\_embedding}.

\begin{figure}
    \centering
    \includegraphics[width=\columnwidth]{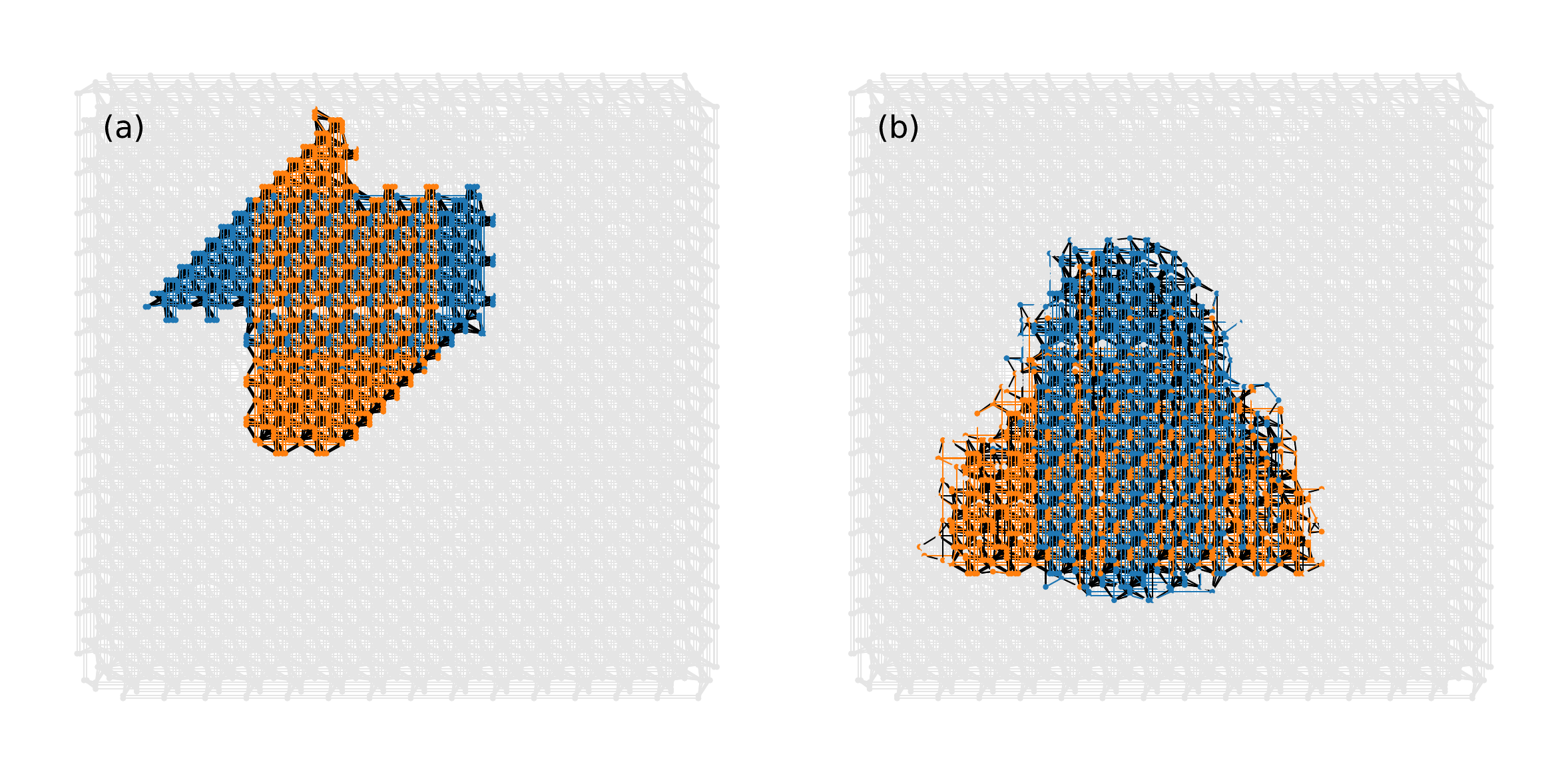}
    \caption{(a) A minor embedding produced by the \texttt{find\_clique\_embedding} function, mapping the single-quarter promotion cannibalization problem onto the hardware graph of the D-Wave Advantage. The quadratic penalty method was used to encode the problem's constraint, resulting in a fully connected logical graph. Nodes and edges, representing physical qubits and couplings, are coloured according to the lowest energy solution sampled by the quantum annealer. Chains of qubits corresponding to logical variables equal to 0 (1) are coloured blue (orange). Edges representing couplings between two different logical variables are coloured black. Unused qubits and couplings are shown in grey. (b) A similar diagram for the case where the \texttt{find\_embedding} function was instead used to calculate the minor embedding. Note that \texttt{find\_embedding} produces a different embedding depending on the value of its seed parameter, so this is an example with one particular parameter value.}
    \label{fig:minor_embedding_heuristics_comparison}
\end{figure}

It appears that the reason for the better average performance of \texttt{find\_embedding} is to do with the fact that the chain lengths in these minor embeddings are so long that the placement of a chain of qubits on the hardware graph strongly influences its value at the end of an anneal. Fig.~\ref{fig:minor_embedding_heuristics_comparison} compares two example minor embeddings using these functions. The chains are coloured based on the values of the variables in the lowest energy solution that was sampled in each case. We can see that in both solutions, roughly half of the chains correspond to logical variables equal to 0 and half equal to 1. This indicates that the problem's constraint of placing half of the products on promotion is close to being satisfied. Looking at the chains corresponding to variables with the same value, we can see a structure that is highly dependent on the locations of the chains. In particular, it appears that the solutions partition the variables in such a way that chains with more physical couplings between them are more likely to take the same logical value. The objective function of the problem does not exhibit such a structure. Therefore, the particular mapping of logical variables to physical qubits must be influencing the values the variables take in the solutions.

For the two examples shown in Fig.~\ref{fig:minor_embedding_heuristics_comparison}, the lowest energy solution that was sampled using the \texttt{find\_embedding} function has a lower energy than the lowest energy solution using \texttt{find\_clique\_embedding}, even though more physical qubits are used with the \texttt{find\_embedding} embedding.
This is because the \texttt{find\_embedding} function happened to result in a more favourable arrangement of variables in the minor embedding when measuring the energy of the best found solution. We note that we used a different value of the seed parameter in \texttt{find\_embedding} for each problem instance, resulting in a variety of different minor embeddings. The example in Fig.~\ref{fig:minor_embedding_heuristics_comparison}(b) is for one particular instance. In comparison, \texttt{find\_clique\_embedding} does not have a seed parameter, so the function produces the same minor embedding for every instance of the problem.

\section{\label{app:1x100_quadratic_penalty_strength_choice}Quadratic penalty strength choice for the single-quarter promotion cannibalization problem}

\begin{figure}
    \centering
    \includegraphics[width=\columnwidth]{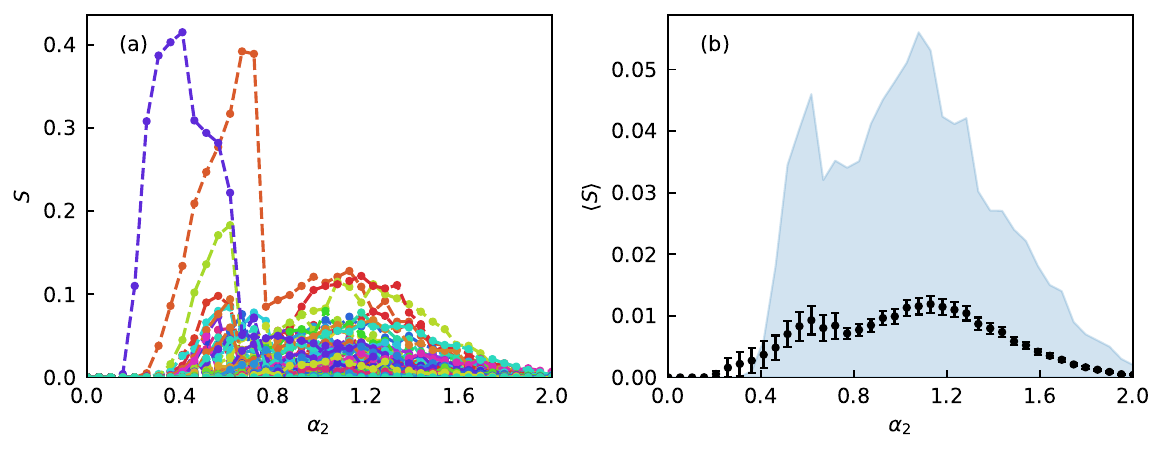}
    \caption{(a) For a simulated annealing algorithm solving the single-quarter promotion cannibalization problem, we plot the fraction $S$ of sampled solutions that are optimal against the quadratic penalty strength $\alpha_2$ for 200 different problem instances, shown in different colours. Lines connecting the points for each instance are shown to guide the eye. (b) We plot the average of $S$ over the 200 instances with error bars representing the standard error in the mean. The blue shaded region contains the 5th to 95th percentile values of $S$.
    }
    \label{fig:1x100_quadratic_penalty_strength_unnormalised}
\end{figure}

The value of the quadratic penalty strength $\alpha_2=1.2$ for our D-Wave experiments on the single-quarter promotion cannibalization problem was chosen based on the results of simulated annealing runs on 200 randomly selected problem instances. One of the metrics we considered was the fraction $S$ of simulated annealing samples that were optimal solutions. This metric was normalised by the largest measured value of $S$ for each problem instance in order to avoid instances with large values of $S$ dominating the average. The average of this normalised data is plotted in Fig.~\ref{fig:1x100_quadratic_penalty_strength}(a). Our choice of $\alpha_2$ was also based on the fraction $F$ of sampled solutions that were feasible, which is plotted in Fig.~\ref{fig:1x100_quadratic_penalty_strength}(b). For transparency, we plot $S$ without normalisation in Fig.~\ref{fig:1x100_quadratic_penalty_strength_unnormalised}(a) for the same selection of problem instances. We also plot the average of this unnormalised data in Fig.~\ref{fig:1x100_quadratic_penalty_strength_unnormalised}(b), which peaks at a similar value as in Fig.~\ref{fig:1x100_quadratic_penalty_strength}(a). The same process was used to determine the quadratic penalty strengths for the four-quarter problem, which is discussed in \ref{app:4x10_quadratic_penalty_strength_choices}.

\section{\label{app:4x10_quadratic_penalty_strength_choices}Quadratic penalty strength choices for the four-quarter promotion cannibalization problem}

\begin{figure*}
    \centering
    \includegraphics[width=\textwidth]{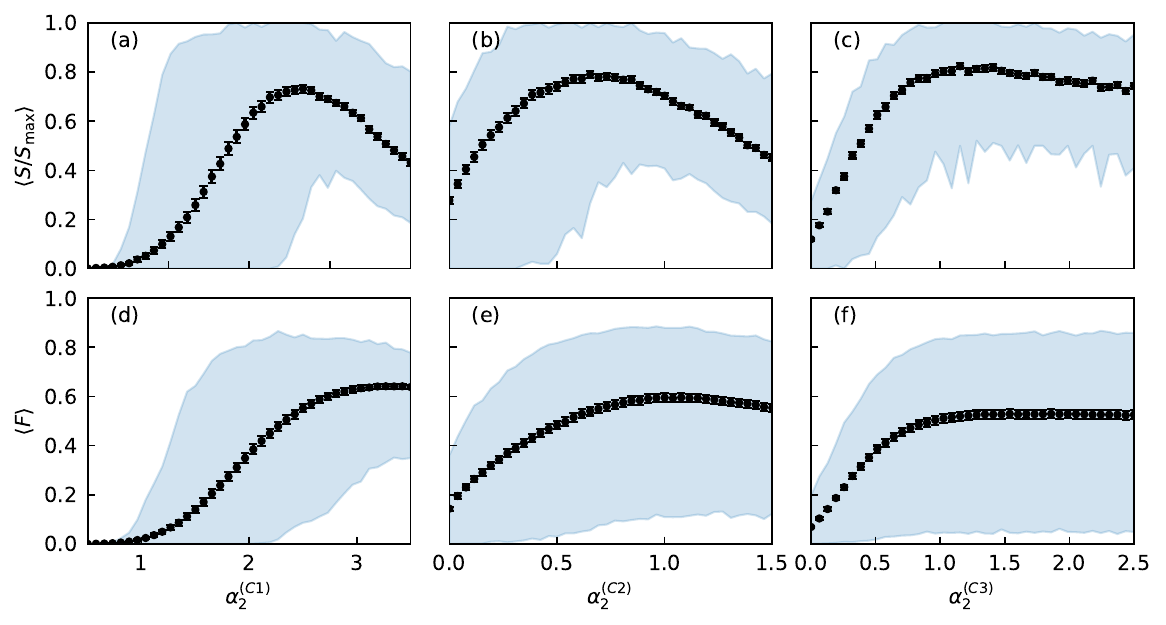}
    \caption{For a simulated annealing algorithm solving the four-quarter promotion cannibalization problem, we have measured the fraction $S$ of sampled solutions that are optimal and normalised this by the maximum measured fraction $S_\mathrm{max}$ for 200 random problem instances. This problem is formulated with three sets of quadratic penalties. In each plot, the penalty strength for one of these sets of penalties is varied along the x-axis while the others are kept constant. The average of $S/S_\mathrm{max}$ is plotted against the quadratic penalty strength for the constraints (a) \ref{constraint:C1}, (b) \ref{constraint:C2}, and (c) \ref{constraint:C3}. Note that $S_\mathrm{max}$ is calculated for each instance separately and is a maximum over the plotted values of $\alpha_2^{(Ci)}$. In (d), (e), and (f), we plot the average fraction $F$ of sampled solutions that are feasible against the quadratic penalty strength for the constraints \ref{constraint:C1}, \ref{constraint:C2}, and \ref{constraint:C3} respectively. The blue shaded regions contain the 5th to 95th percentile values of $S/S_\mathrm{max}$ or $F$, and error bars represent the standard errors in the means. Where kept constant, the quadratic penalty strengths are $\alpha_2^{(C1)} = 2.4$, $\alpha_2^{(C2)} = 0.6$, and $\alpha_2^{(C3)} = 1.2$.}
    \label{fig:4x10_quadratic_penalty_strengths}
\end{figure*}

The values of the quadratic penalty strengths used in the four-quarter promotion cannibalization problem experiments on the D-Wave annealer were chosen based on the results of a simulated annealing algorithm, as was done for the single-quarter problem experiments. For the four-quarter problem, we assigned each of the three sets of constraints \ref{constraint:C1}, \ref{constraint:C2}, \ref{constraint:C3} a separate quadratic penalty strength, which we denote $\alpha_2^{(C1)}$, $\alpha_2^{(C2)}$, and $\alpha_2^{(C3)}$ respectively. The penalty strengths that we picked are $\alpha_2^{(C1)} = 2.4$, $\alpha_2^{(C2)} = 0.6$, and $\alpha_2^{(C3)} = 1.2$. The values were chosen to produce large probabilities of sampling optimal and feasible solutions based on the simulated annealing results shown in Fig.~\ref{fig:4x10_quadratic_penalty_strengths}. 

\section{\label{app:linear_penalty_search_strategy}Linear penalty strength search strategy for multiple linear penalties}

When applying the linear penalty method to more than one constraint in a problem, the process of searching for values of the penalty strengths $\alpha_1$ is more complex than when there is only a single linear penalty. This is because changing a penalty strength for one constraint can affect the assignment of variables in another constraint. To determine which four-quarter promotion cannibalization problem instances could have all four constraints~\ref{constraint:C1} satisfied with linear penalties, we used an $\alpha_1$ search strategy that is outlined with pseudocode in Algorithm~\ref{alg:alpha_1_search}. The function \texttt{linear\_penalty\_strength\_search} attempts to find values of $\alpha_1$ such that Eq.~\ref{eq:constraint_C1} is satisfied for each fiscal quarter, where $A$ is the constraint value. It makes use of an exact solver that is called through the $get\_num\_ones$ function, which returns the Hamming weight of each fiscal quarter in the optimal solution. If \texttt{linear\_penalty\_strength\_search} finds the desired values of $\alpha_1$ within the maximum number of iterations specified by the parameters $max\_iterations\_1$ and $max\_iterations\_2$, it returns these values. If not, it returns $Null$. In principle, a search strategy with a similar structure can be used with a solver that is not exact, such as a quantum annealer.

\begin{algorithm}
\scriptsize
\caption{Four-quarter linear penalty strength search.}\label{alg:alpha_1_search}
\begin{algorithmic}[1]
\Function{linear\_penalty\_strength\_search}{$A$, $max\_iterations\_1$, $max\_iterations\_2$}
    \State $target\_num\_ones \gets [A, A, A, A]$ \Comment{Target Hamming weight for each fiscal quarter}
    \State $pen\_strengths \gets [-1, -1, -1, -1]$ \Comment{Initial linear penalty strengths for each quarter}
    \State $prev\_pen\_strengths \gets [0, 0, 0, 0]$ \Comment{Array to store previous penalty strengths}
    \State $step \gets 0.5$ \Comment{Penalty strength step size}
    \State $steps \gets [step, step, step, step]$
    \State $num\_iterations \gets 0$
    \State $average\_converged \gets False$ \Comment{Flag for convergence of the average Hamming weight}
    \State $high, low \gets Null, Null$ \Comment{Variables to store penalty strengths that are too large or small}
    \While{$(\NOT~average\_converged)~\AND~num_\_iterations<max\_iterations\_1$}
        \State $num\_iterations \gets num\_iterations + 1$
        \State $quarters\_num\_ones \gets get\_num\_ones(pen\_strengths)$ \Comment{Call solver and get an array of Hamming weights}
        \State $av\_num\_ones \gets mean(quarters\_num\_ones)$ \Comment{Average over the Hamming weights for each quarter}
        \If{$av\_num\_ones < A$} \Comment{If the average Hamming weight is too small}
            \State $high \gets pen\_strengths[0]$ \Comment{Penalty strength was too high}
            \If{$low = Null$} \Comment{If a penalty strength that is too low hasn't been found yet}
                \State $prev\_pen\_strengths \gets pen\_strengths$
                \State $pen\_strengths \gets pen\_strengths - steps$ \Comment{Reduce the penalty strength by the step size}
            \Else \Comment{Otherwise, perform a binary search}
                \State $prev\_pen\_strengths \gets pen\_strengths$
                \State $pen\_strengths \gets 0.5 \times (low + high)$
            \EndIf
        \ElsIf{$av\_num\_ones > A$} \Comment{If the average Hamming weight is too large}
            \State $low \gets pen\_strengths[0]$ \Comment{Penalty strength was too low}
            \If{$high = Null$} \Comment{If a penalty strength that is too high hasn't been found yet}
                \State $prev\_pen\_strengths \gets pen\_strengths$
                \State $pen\_strengths \gets pen\_strengths + steps$ \Comment{Increase the penalty strength by the step size}
            \Else \Comment{Otherwise, perform a binary search}
                \State $prev\_pen\_strengths \gets pen\_strengths$
                \State $pen\_strengths \gets 0.5 \times (low + high)$
            \EndIf
        \Else
            \State $average\_converged \gets True$
        \EndIf
    \EndWhile
    \If{$quarters\_num\_ones = target\_num\_ones$}
        \State \Return $pen\_strengths$ \Comment{If all quarters' Hamming weights converged to $A$, we are done}
    \EndIf
    \State $num\_iterations \gets 0$
    \State $converged\_all = False$ \Comment{Flag for convergence of all quarters' Hamming weights}
    \State $step \gets 0.1 \times step$ \Comment{Reduce the step size before the second search}
    \While{$(\NOT~converged\_all)~\AND~num\_iterations < max\_iterations\_2$}
        \State $num\_iterations \gets num\_iterations + 1$
        \State $quarters\_num\_ones \gets get\_num\_ones(pen\_strengths)$ \Comment{Call solver and get an array of Hamming weights}
        \State $converged\_all = True$
        \For{$q \gets$ from $0$ to $3$} \Comment{For each fiscal quarter}
            \If{$quarters\_num\_ones[q] < A$} \Comment{If the Hamming weight for the quarter is too low}
                \State $converged\_all \gets False$
                \If{$pen\_strengths[q] > prev\_pen\_strengths[q]$}
                    \State $new\_pen\_strength \gets 0.5 \times (prev\_pen\_strengths[q] + pen\_strengths[q])$
                \Else
                    \State $new\_pen\_strength \gets pen\_strengths[q] - step$
                \EndIf
                \State $prev\_pen\_strengths[q] \gets pen\_strengths[q]$
                \State $pen\_strengths[q] \gets new\_pen\_strength$
            \ElsIf{$quarters\_num\_ones[q] > A$} \Comment{If the Hamming weight for the quarter is too high}
                \State $converged\_all \gets False$
                \If{$pen\_strengths[q] < prev\_pen\_strengths[q]$}
                    \State $new\_pen\_strength \gets 0.5 \times (prev\_pen\_strengths[q] + pen\_strengths[q])$
                \Else
                    \State $new\_pen\_strength \gets pen\_strengths[q] + step$
                \EndIf
                \State $prev\_pen\_strengths[q] \gets pen\_strengths[q]$
                \State $pen\_strengths[q] \gets new\_pen\_strength$
            \EndIf
        \EndFor
        \State $step = step \times 0.93$ \Comment{Reduce the step size in each iteration}
    \EndWhile
    \If{$\NOT~converged\_all$}
        \State \Return $Null$ \Comment{Return null if the search failed}
    \EndIf
    \State \Return $pen\_strengths$
\EndFunction
\end{algorithmic}
\end{algorithm}

The strategy performs two searches. The first search, which is performed in lines 10---33 in Algorithm~\ref{alg:alpha_1_search}, uses the same value of $\alpha_1$ for each constraint and attempts to tune the value such that the average Hamming weight of the variables of the four quarters is equal to $A$. It performs a maximum of $max\_iterations\_1$ iterations, where in each iteration, $\alpha_1$ is increased if the average Hamming weight is larger than $A$ or decreased if it is greater than $A$. The value of $\alpha_1$ is increased or decreased by an amount given by the step size $step$ until a value that is too low and one that is too high is found. After that, a binary search is used to narrow down on a value of $\alpha_1$ for which the average constraint value is equal to $A$ in the optimal solution, if it exists. Sometimes, this first search finds a value of $\alpha_1$ where not only does the average constraint value equal $A$, but also each quarter satisfies its individual constraint. In this case, the search is complete.

In cases where the first search does not produce an optimal solution in which all four constraints are satisfied, a second search is performed. This corresponds to lines 37---58 in Algorithm~\ref{alg:alpha_1_search}. This second search performs a maximum of $max\_iterations\_2$ iterations with a smaller step size $step$ than in the first search. The key difference is that in the second search, the values of $\alpha_1$ for the four quarters are tuned individually. For each quarter, $\alpha_1$ is increased or decreased depending on the Hamming weight of the variables associated with that particular quarter. If the direction in which $\alpha_1$ is being changed is opposite to the direction of the previous step, then the value of $\alpha_1$ is set to the midpoint of its current and previous values. Otherwise, the amount by which $\alpha_1$ is changed is equal to $step$. In each iteration, the value of $step$ is reduced so that the search becomes increasingly more precise.

In our runs of this search strategy, we used the parameter values $max\_iterations\_1 = 20$ and $max\_iterations\_2 = 100$. However, the average number of iterations performed by the search strategy was $\approx 13$ for problem instances that we were able to find good $\alpha_1$ values for, implying that much smaller values of $max\_iterations\_1$ and $max\_iterations\_2$ can be used for this problem. Furthermore, the number of iterations could be reduced by optimizing the values of $step$ on line 5 of Algorithm~\ref{alg:alpha_1_search} and the multiplicative factors on lines 36 and 58.

In parts of our work, we considered penalty schemes where some of the constraints~\ref{constraint:C1} are implemented with linear penalties and others with quadratic penalties. The $\alpha_1$ search strategy in Algorithm~\ref{alg:alpha_1_search} was also used in these cases. To account for the quadratic penalties, the function $get\_num\_ones$ was changed to apply quadratic penalties instead of linear penalties to the relevant quarters. The quarters with quadratic penalties applied always have the correct Hamming weight in the optimal solution as long as $\alpha_2$ is chosen to be large enough. The function \texttt{linear\_penalty\_strength\_search} returns linear penalty strengths even for quarters that have quadratic penalties applied, which can be ignored.

\end{document}